\documentclass[article,amsmath,amssymb,floatfix,aps]{revtex4}
\UseRawInputEncoding
\usepackage{graphicx}
\usepackage{epsfig}
\usepackage{multirow}
\usepackage{color}
\usepackage{cancel}
\begin{document}

\title{A new boson approach for the wobbling motion in even-odd nuclei}

\author{A. A. Raduta$^{a), b)}$,  C. M. Raduta $^{a)}$ and  R. Poenaru $^{a), c)}$}

\affiliation{$^{a)}$ Department of Theoretical Physics, Institute of Physics and
  Nuclear Engineering, Bucharest, POBox MG6, Romania}

\affiliation{$^{b)}$Academy of Romanian Scientists, 54 Splaiul Independentei, Bucharest 050094, Romania}

\affiliation{$^{c)}$Doctoral School of Physics, Bucharest University, 405 Atomistilor Str., Bucharest-Magurele, Romania}

\begin{abstract}
A triaxial core rotating around the middle axis, i.e. 2-axis, is cranked around the  1-axis, due to the coupling of an odd proton from a high j orbital. 
Using the Bargmann representation of a new and complex boson expansion of the angular momentum components, the eigenvalue equation of the model Hamiltonian
 acquires a Schr\"{o}dinger form with a fully separated  kinetic energy. From a critical angular momentum, the potential energy term exhibits three minima, two of them being degenerate. Spectra of the deepest wells reflects a chiral-like structure. Energies corresponding to the deepest and local minima respectively,  are analytically expressed within a harmonic approximation. Based on a classical analysis, a phase diagram is constructed. It is also shown that the transverse wobbling  mode is unstable.  The wobbling frequencies corresponding to the deepest minimum are used to quantitatively describe the wobbling properties in $^{135}$Pr. Both energies and e.m. transition probabilities are described.
\end{abstract}
\pacs{21.10.Re, 21.60.Ev,27.70.+q}
\maketitle

\renewcommand{\theequation}{1.\arabic{equation}}
\setcounter{equation}{0}
\section{Introduction}

The wobbling motion is defined as a precession of the total angular momentum combined with an oscillation of its projection on the quantization axis around a steady position.
This phenomenon was first described by Bohr and Mottelson within a triaxial rotor  model for high spin states, where the total angular momentum almost aligns to the principal axis with the largest moment of inertia \cite{BMott}.  A fully microscopic formalism,  due to Marshalek \cite{Marsh}, followed the mentioned pioneering paper. Within the passed  time interval, a large volume of experimental and theoretical results has been reported
\cite{Odeg,Jens,Ikuko,Scho,Amro,Gorg,Ham,Matsu,Ham1,Jens1,Hage,Tana3,Oi,Rad07,Bring,Hage1,Hart,Cast,Alme,MikIans,Rad016,Badea,Rad017,Rad018,Rad20,Rad201}. Also, the extension of the concept to the even-odd nuclei showed up. The excited wobbling states  are known in several even-odd triaxial nuclei like $^{161,163,165,167}$Lu 
\cite{Jens,Scho,Amro,Hage,Bring,Hage1}, $^{167}$Ta \cite{Hart}, $^{135}$Pr  
\cite{Tan017,Tana018,Frau,Frau018,Buda,Matta,Sen}, $^{187}$Au \cite{Sen1}, $^{133}$La\cite{Biswas}, $^{105}$Pd \cite{Timar} and $^{133}$Ba \cite{Devi}.

With the time, several formalisms were attempted to describe the wobbling motion in nuclei. Thus, the classical interpretation of Bohr and Mottelson was widely used by various authors in the context of interpreting the new data that meanwhile appeared \cite{Ikuko,Ham,Frau,Chen1,Lawr}. The oldest and simplest boson description of the wobbling phenomenon belongs to Bohr and Mottelson
\cite{BMott}. More elaborate interpretations are those of Refs. \cite{Tana3,Oi,Badea}, where Holstein-Primakoff \cite{HolPr} and Dyson \cite{Dys} boson expansions were used, respectively. The semi-classical studies
\cite{Rad07,Rad017,Rad018,Buda,Rad20,Rad201} proved to be an efficient and flexible tool for a realistic view of this phenomenon in even-odd nuclei. 

The wobbling states are actually fingerprints of the triaxial structure of the nuclei, which justifies the attractive appeal of the subject. The first paper devoted to triaxial nuclei was that of Davydov and Fillipov \cite{Davy}. The $\gamma$ deformation of the atomic nuclei has been treated by many authors \cite{WilJean,TerVehn,Bona2,Bona3,Rad09}.  In Ref.\cite{Davyd}, Davydov introduced, for the first time in the literature, an  appropriate form of the Hamiltonian for an even-odd nucleus consisting in a core and an odd particle moving in a potential, coupling it to the collective core. This Hamiltonian is nowadays  widely used by theoreticians. The first results reported for even-odd nuclei within a quasiparticle plus triaxial rotor framework, for the rare earth region, were given in Refs. \cite{Toki1,Toki2,Toki3,Toki4,Toki5}. 

The wobbling motion has a longitudinal/transverse character depending on whether  the relative position of the odd-particle angular momentum and the core axis with the largest MoI are parallel/perpendicular 
\cite{Frau}. Although the concept of transverse wobbling is being used by many authors, a certain debate on whether such a wobbling motion exists or not is still standing \cite{Rad201,Tana018,Frau018,Lawr}.

In the present paper we propose a new formalism based on a boson expansion of the angular momentum components. A particular case of the new expansion is a generalization of the Dyson boson representation. By using the Bargmann representation \cite{Bar}, the eigenvalue equation for the model Hamiltonian is brought to a Schr\"{o}dinger form which, in the harmonic approximation, leads to an explicit expression for the wobbling frequency. Also, a semi-classical description is provided, which is fully consistent with the quantal treatment. Within this picture, it is proved that the transversal
mode is unstable despite the fact that we assumed that the middle axis is of maximal MoI. The formalism is applied, with a positive result, to $^{135}$Pr.

The project sketched above is accomplished according to the following plan. Using a new boson expansion for the angular momentum, in section II the Schr\"{o}dinger equation for the model Hamiltonian is derived, while in section III, another boson expansion is obtained. The harmonic approximation is delivered in Section IV, while the classical approach is presented in Section V.
In Section VI we  give the analytical formulas for the electromagnetic reduced transition probabilities. Numerical results and discussions are presented in Section VII, while the summary and the final conclusions are described in section VIII.

\renewcommand{\theequation}{2.\arabic{equation}}
\setcounter{equation}{0}
\section{A compact formula for the potential energy of a particle-triaxial rotor Hamiltonian}
Assuming a rigid coupling of an odd nucleon to a triaxial core, the Hamiltonian for the even-odd system may be approximated as:
\begin{equation}
\hat{H}_{rot}=\sum_{k=1,2,3}A_k(\hat{I}_k-\hat{j}_k)^2,
\end{equation}
with $A_k=\frac{1}{2{\cal J}_k}$  and $I$ standing for the total angular momentum.

For a rigid coupling of the odd proton to the triaxial core, we suppose that $\bf{j}$ stays in the principal plane (1,2): ${\bf j}=(j\cos\theta, j\sin \theta, 0)$.
Also,  we consider that the maximal moment of inertia (MoI) is ${\cal J}_2$; furthermore, we expand the linear term in $I_2$ as the first order of approximation:
\begin{equation}
\hat{I}_2=I\left(1-\frac{1}{2}\frac{\hat{I}_1^2+\hat{I}_3^2}{I^2}\right).
\end{equation}
Thus, the Hamiltonian acquires the form:
\begin{equation}
\hat{H}_{rot}=A\hat{H}^{\prime}+(A_1I^2-A_2j_2I)+\sum_{k=1,2}A_k\hat{j}_k^2,
\label{hahhasprim}
\end{equation}
where the following notations have been used:
\begin{eqnarray}
\hat{H}^{\prime}&=&\hat{I}_2^2+u\hat{I}_3^2+2v_0\hat{I}_1,\;\;\rm{with}\nonumber\\
u&=&\frac{A_3-A_1}{A},\;\;v_0=\frac{-A_1j_1}{A},\;\;A=A_2(1-\frac{j_2}{I})-A_1
\end{eqnarray}
In order that the classical Hamiltonians associated with $\hat{H}_{rot}$ and $\hat{H}^{\prime}$ respectively, exhibit the same stationary points, it is necessary that $A>0$.
For what follows, we suppose that the MoI's are such that $1>u>-1$.

A Hamiltonian similar to $\hat{H}^{\prime}$, but describing an even-even nucleus, was studied in both semi-classical and  quantal frameworks in Ref. \cite{Rad98}. Here we focus our attention on the quantal description. Note that $\hat{H}^{\prime}$ looks like a Hamiltonian for a triaxial rotor amended with a new term, which cranks the system to rotate around the one-axis. It is convenient to choose  the cranking axis as quantization axis. Moreover, it is useful to express the considered Hamiltonian in terms of the raising and lowering angular momenta operators:
\begin{equation}
\hat{I}_{\pm}=\hat{I}_2\pm i\hat{I}_3,\;\;\hat{I}_0=\hat {I}_1.
\end {equation}
In the rotating frame of reference, the angular momentum components satisfy the commutation relations:
\begin{equation}
\left[\hat{I}_{-},\hat{I}_{+}\right]=2\hat{I}_{0},\;\;\left[\hat{I}_{\mp},\hat{I}_{0}\right]=\mp\hat{I}_{\mp}.
\end{equation}
In terms of the new variables, one obtains:
\begin{equation}
\hat{H}^{\prime}=\frac{1-u}{4}\left(\hat{I}_{+}^{2}+\hat{I}_{-}^{2}\right)+\frac{1+u}{4}\left(\hat{I}_{+}\hat{I}_{-}+\hat{I}_{-}\hat{I}_{+}\right)+2v_0\hat{I}_0.
\end{equation}
The Schr\"{o}dinger equation associated with $\hat {H}^{\prime}$,
\begin{equation} 
\hat {H}^{\prime}|\Psi\rangle =E|\Psi\rangle,
\end{equation}
is further written in terms of the conjugate variables $q$ and $\frac{d}{dq}$, by using the following representation for the angular momentum components:

\begin{equation}
\hat{I}_{\mp}=i\frac{c\pm d}{k's}\left(I\mp \hat{I}_0\right),\;\;
\hat{I}_0=Icd-s\frac{d}{dq}\equiv \hat{I}_1,
\label{amqdq}
\end{equation} 
where $s, c$ and $d$ denote the Jacobi elliptic functions:
\begin{eqnarray}
s&=&sn(q,k),\;\;c=cn(q,k),\;\;d=dn(q,k),\;\;\rm{with}\nonumber\\
k&=&\sqrt{|u|},\;\;k^{\prime}=\sqrt{1-k^2},\nonumber\\
q&=&\int_{0}^{\varphi}\left(1-k^2\sin^2(t)\right)^{-1/2}dt\equiv F(\varphi,k).
\end{eqnarray}
The dependence of the Jacobi functions on the variable $q$ is shown in Fig.1.
Their connection with the trigonometric function is given by:
\begin{equation}
s=\sin\varphi,\;\;c=\cos \varphi,\;\;d=\sqrt{1-k^2s^2}.
\end{equation}

\begin{figure}[ht!]
\begin{center}
\includegraphics[width=0.5\textwidth]{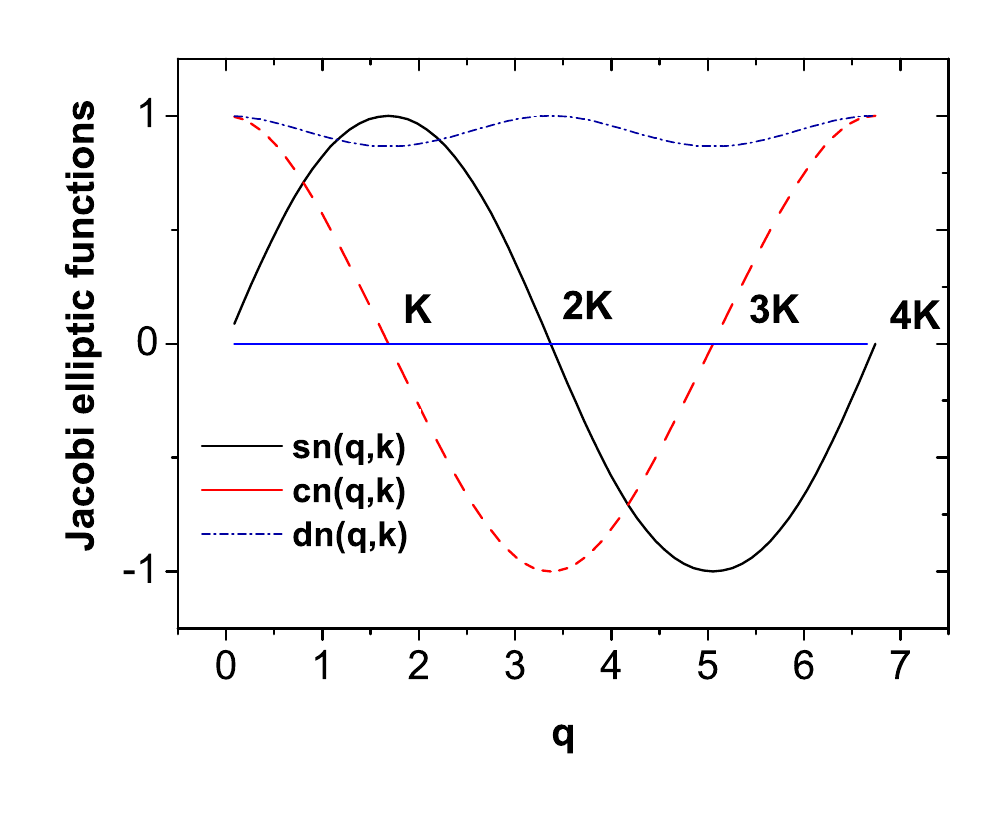}
\end{center}
\caption{(Color online) The elliptic functions sn, cn, and dn are represented as function of $q$, for $k=1/2$.}
\label{Fig.1}
\end{figure}

Obviously, the functions $s, c$ and $d$ are periodic  in $q$, with the periods
$4K, 4K$ and $2K$ respectively, where:
\begin{equation}
K=\frac{\pi}{2} {{_2}F}{_1}(\frac{1}{2},\frac{1}{2},1;k^2).
\label{Ka}
\end{equation}

\begin{figure}
\includegraphics[width = 0.5\textwidth]{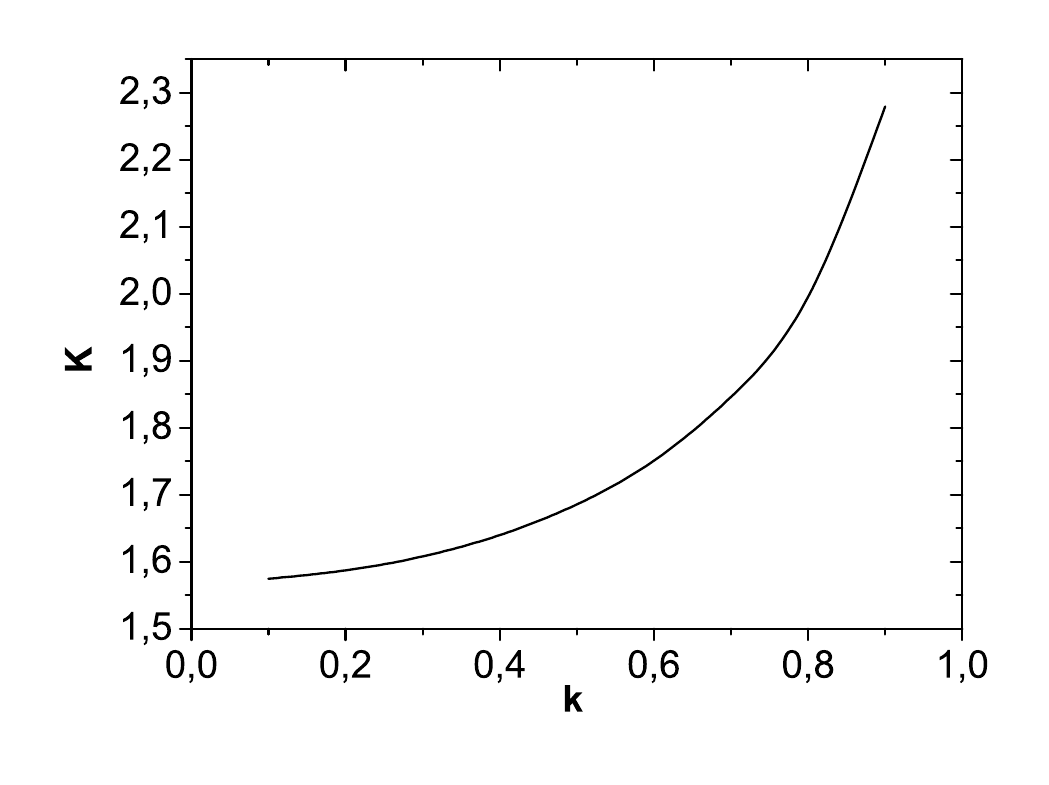}
\caption{(Color online) The period $K$ given by Eq.\ref{Ka} is plotted as function of $k$.}
\label{Fig.2}
\end{figure}

The standard notation for the hyper-geometric function $_2F_1(\alpha,\beta,\gamma;\epsilon)$, has been used. The magnitude $K$ is plotted as function of $k$ in Fig.2.
In terms of the newly introduced conjugate coordinates, $\hat{H}^{\prime}$ becomes:
\begin{equation}
‍\hat{H}^{\prime}=-\frac{d^2}{dq^2}-2v_0s\frac{d}{dq}+I(I+1)s^2k^2+2v_0cdI.
\label{Hqdq}
\end{equation}
Changing the wave-function by the transformation:
\begin{equation}
|\Psi\rangle=(d-kc)^{-\frac{v_0}{k}}|\Phi\rangle,
\end{equation}
the Schr\"{o}dinger equation acquires a new form, where the kinetic and potential energies are separated:
\begin{equation}
\left[-\frac{d^2}{dq^2}+V(q)\right]|\Phi\rangle =E|\Phi \rangle .
\label{Scheq}
\end{equation}
The potential energy term has the expression:
\begin{equation}
V(q)=\left[I(I+1)k^2+v_0^2\right]s^2+(2I+1)v_0cd.
\end{equation}

\begin{figure}
\includegraphics[width=0.5\textwidth]{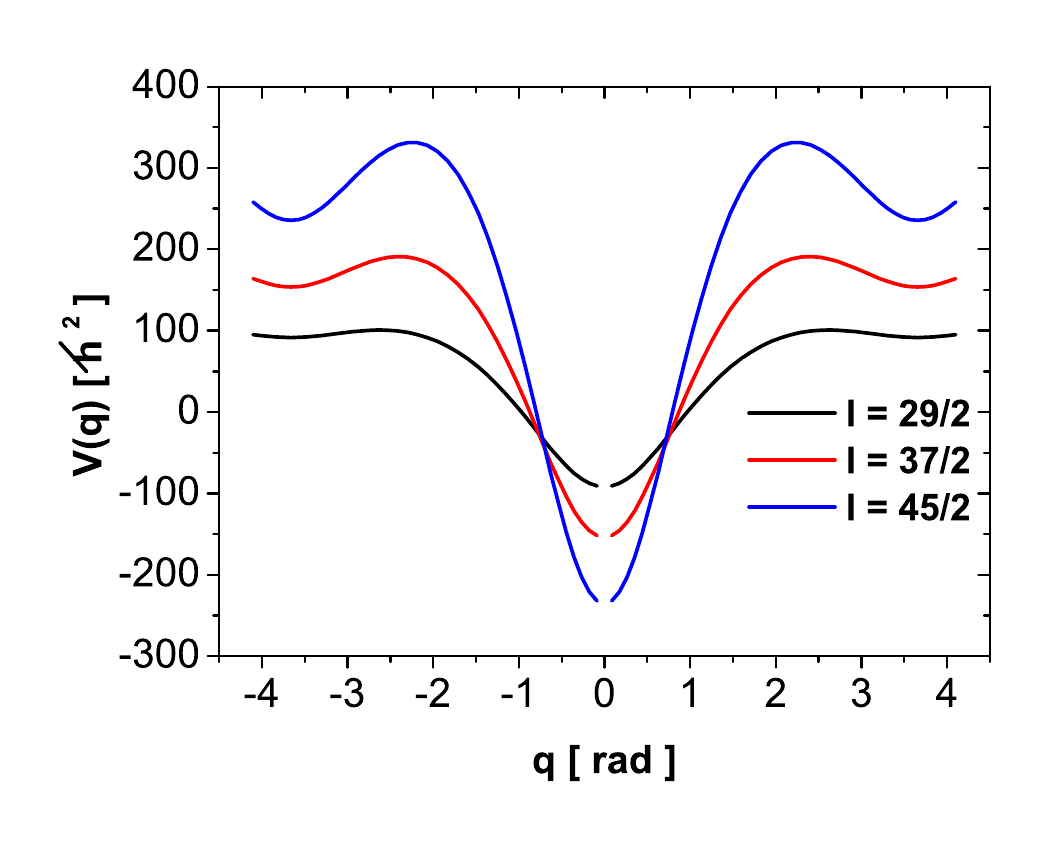}
\caption{(Color online) The potential energy is plotted as function of $q$ for a particular set of values for the moment of inertia ($MoI$) 
: ${\cal J}_1:{\cal J}_2:{\cal J}_3=95:100:85\hbar^2 MeV^{-1}$,the odd particle angular momentum j=13/2 and $\theta = - 80^{0} $.}
\label{Fig.3}
\end{figure}
It is worth mentioning that the transformation (\ref{amqdq}) preserves the commutation relations obeyed by the angular momentum components (2.6).

The shape of the potential energy term is shown in Fig. 3. Note that V(q) is invariant with respect to the transformation $q\to-q$. This leads to the fact that in the interval, for example of
 [-4K,4K], the potential exhibits three deep symmetric wells with  degenerate minima, and two local minima  in $q=\pm 2K$. States inside the local minima are meta-stable, since they are tunnelling to the adjacent deep minima. The states in the deepest wells are degenerate.
The shape of the potential V(q) in the interval of [-2K,2K] is shown in Fig.3, for a few angular momenta I. To visualize the symmetry mentioned above, we plotted V(q) in a larger interval, namely [-4K,4K], for $I=45/2$ and $\theta =-80^0$ (Fig. 4) and $\theta=100^0$ (Fig. 5), respectively.

\begin{figure}
\includegraphics[width=0.4\textwidth]{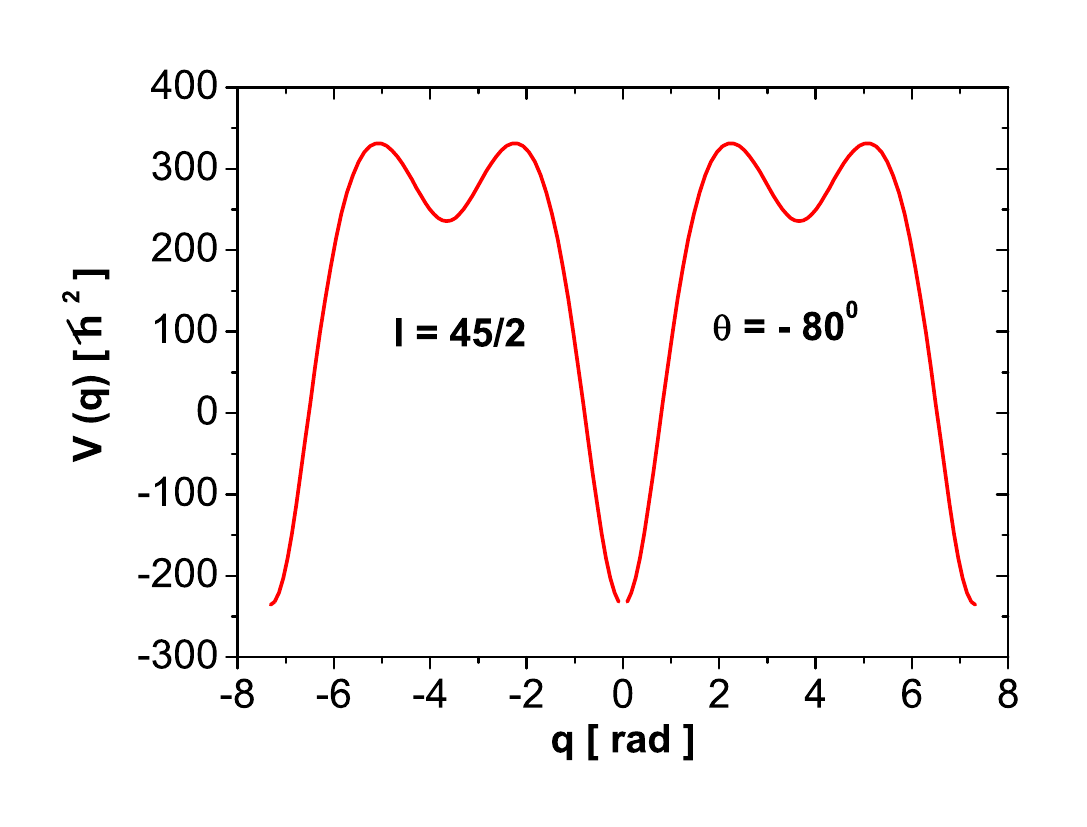}
\caption{(Color online) The potential energy is plotted as function of $q$ for a particular set of values for the moment of inertia ($MoI$) 
: ${\cal J}_1:{\cal J}_2:{\cal J}_3=95:100:85\hbar^2 MeV^{-1}$ and $\theta = - 80^{0} $. Negative values for q are also included. The total angular momentum is I=45/2 and j=13/2.}
\label{Fig.4}
\end{figure}
\begin{figure}
\includegraphics[width=0.4\textwidth]{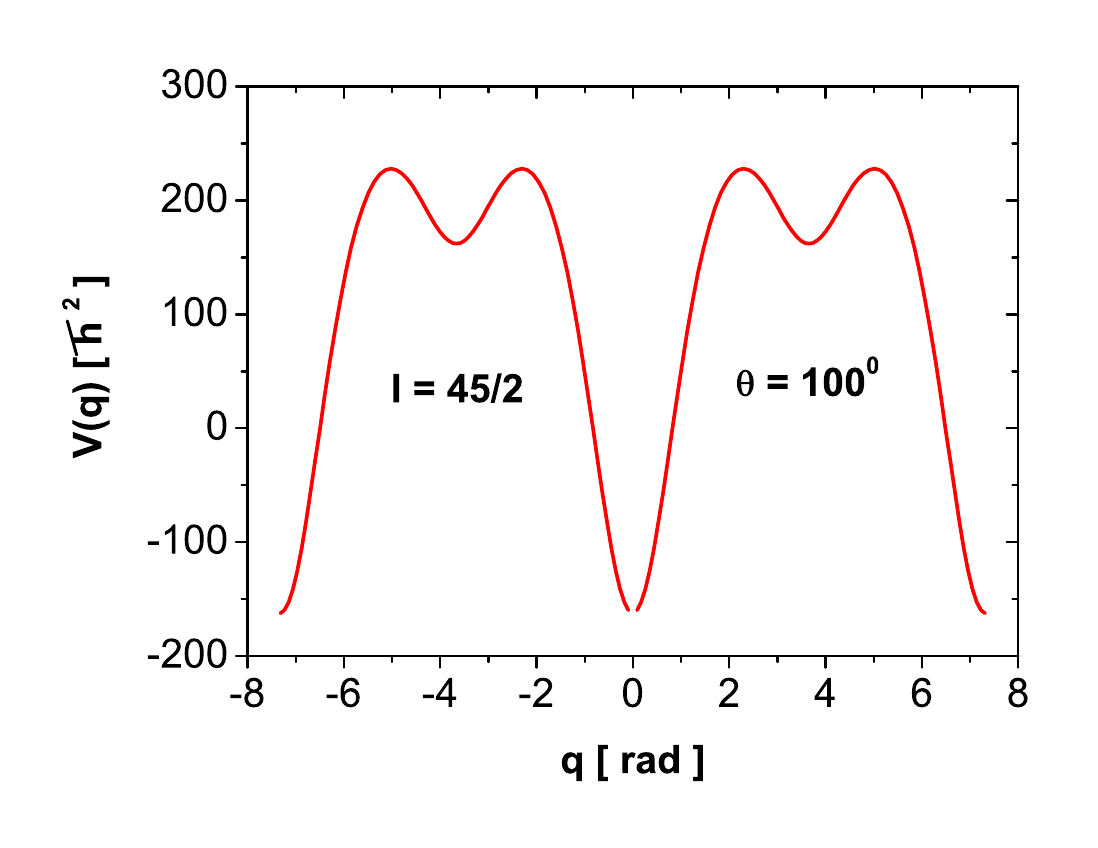}
\caption{(Color online) The potential energy is plotted as function of $q$ for a particular set of  values for the moment of inertia ($MoI$) 
: ${\cal J}_1:{\cal J}_2:{\cal J}_3=95:100:85\hbar^2 MeV^{-1}$, and $\theta = 100^{0} $. Negative values for q are also included. The total angular momentum is I=45/2 and j=13/2.}
\label{Fig.5}
\end{figure}

In order to prove that the transformation (2.9) preserves the commutation relations for the angular momentum components, we need the first derivatives of the Jacobi functions:
\begin{equation}
\frac{d}{dq}sn(q)=cn(q)dn(q),\;\;
\frac{d}{dq}cn(q)=-sn(q)dn(q),\;\;
\frac{d}{dq}dn(q)=-k^2sn(q)cn(q).
\end{equation}
Note now that by using the Bargmann \cite{Bar} mapping to the boson operators $b,b^{\dagger}$: 
\begin{equation}
q\to b^{\dagger},\;\;\frac{d}{dq}\to b,
\label{bostr}
\end{equation}
it becomes manifest that Eq. (\ref{amqdq}) expresses a homeomorph mapping of the angular momentum components, i.e. the generators of an SU(2) algebra, onto a  boson realization of the mentioned  algebra. Indeed, within the Bargmann representation we have:

\begin{eqnarray}
\hat{I}_{+}&=&i\frac{c(b^{\dagger})-d(b^{\dagger})}{k^{\prime}s(b^{\dagger})}\left (I+Ic(b^{\dagger})d(b^{\dagger})-s(b^{\dagger})b\right ),\nonumber\\
\hat{I}_{-}&=&i\frac{c(b^{\dagger})+d(b^{\dagger})}{k^{\prime}s(b^{\dagger})}\left (I-Ic(b^{\dagger})d(b^{\dagger})+s(b^{\dagger})b\right ),\nonumber\\
\hat{I}_{0}&=&Ic(b^{\dagger})d(b^{\dagger})-s(b^{\dagger})b.
\label{bosexp}
\end{eqnarray}
To our knowledge this is the first time when such a boson "expansion" shows up in the literature. Obviously, this is different from the known boson expansions proposed by Holstein-Primakoff \cite{HolPr}, and Dyson \cite{Dys}. We note that like the Dyson's, this boson representation does not preserve the hermiticity. Indeed, one can easily check that:
\begin{equation}
\left(\hat{I}_{+}\right)^+\ne \hat{I}_{-}
\end{equation}
Also, the boson Hamiltonian obtained from (\ref{Hqdq}) by using he transformation (\ref{bostr}), is not Hermitian. However, it may be shown \cite{Ogu} that it has real eigenvalues. For our further purposes it is convenient not to use the boson Hamiltonian, but rather the Schr\"{o}dinger equation (\ref{Scheq}). We remark that in order to make the Holstein-Primakoff boson expansion tractable, the involved square root operators must be expanded in power series of 
$\hat{N}/I$, with $\hat{N}$ denoting the boson number operator, while $I$ is the total angular momentum. This expansion is truncated in the second order, and moreover no contribution caused by the normal ordering of the higher order terms, is included. One remarks  the fact that the whole boson series involves higher order terms in the linear momentum, which  conflicts with the semi-classical framework. In contradistinction to such a case the boson Hamiltonian associated with $\hat{H}^{\prime}$ is written in a normal order, and is quadratic in the linear momentum $-i\frac{d}{dq}$.

In case of the Holstein-Primakoff boson expansion, a certain caution concerning Hermiticity and $D_2$ invariance [1], should be payed. In order to overcome the situation when one of these symmetries is broken,  one has to extend the expansion up to the next-to-leading orders in boson operators[14]. Note that the asymmetric treatment emerging from the Bargmann mapping may affect the final results of the numerical analysis.

\renewcommand{\theequation}{3.\arabic{equation}}
\setcounter{equation}{0}
\section{Another new boson expansion for the a.m. components}
The  case of $k=0$ deserves a special  attention. Indeed, for this value of $k$, one obtains:
\begin{equation}
q=\varphi;\;\;,d=1;\;\;K=\frac{\pi}{2},\;\;k^{\prime}=1.
\end{equation}
Using these simple relations in connection with Eq.(\ref{bosexp}), one obtains a new boson expansion for the angular momentum components in the rotating frame:
\begin{eqnarray}
I_+&=&i\left[-I\sin b^{\dagger}+(1-\cos b^{\dagger})b\right],\nonumber\\
I_-&=&i\left[I\sin b^{\dagger}+(1+\cos b^{\dagger})b\right],\nonumber\\
I_0&=&I\cos b^{\dagger}-\left(\sin b^{\dagger}\right)b.
\label{bex2}
\end{eqnarray}
Again, this boson expansion is a particular case of Eq.(\ref{bosexp}), and different from the traditional ones mentioned above, as due to Holstein-Primakoff and Dyson.
Expanding, consistently,   the trigonometric functions, and keeping only the leading terms, we obtain:
\begin{eqnarray}
I_{+}&=&i\left[-Ib^{\dagger}+\frac{1}{2}(b^{\dagger})^2b \right],\nonumber\\
I_{-}&=&2ib,\nonumber\\
I_{0}&=&I-b^{\dagger}b ,
\end{eqnarray}
which is just the Dyson boson expansion of the angular momentum components. Due to this result, we may assert that the  boson expansions given by Eqs.(\ref{bosexp}), and (\ref{bex2})
respectively, represent two distinct generalizations of the well known Dyson boson expansion.

\renewcommand{\theequation}{4.\arabic{equation}}
\setcounter{equation}{0}
\section{Harmonic approximation}
Eq.(\ref{Scheq}) can be numerically solved. However, as we shall further see, there are arguments for the validity of the harmonic approximation.
First, we look for the stationary point of the potential energy term. These are obtained by looking for the roots of the first derivative of $V(q)$:
\begin{eqnarray}
V^{\prime}(q)&=&s\left[\left(I(I+1)k^2+v_0^2\right)2cd-(2I+1)v_0{k^{\prime}}^2\right.\nonumber\\
&-&\left.(2I+1)v_02k^2c^2\right].
\end{eqnarray}
Among the stationary points there are five minima for $q=0,\pm 2K,\pm 4K$, respectively. As shown in Fig. 3, the minima  $q=\pm 2K$ show up only for $I>19/2$. The two minima are flat at the beginning, but their depth increases with the spin. The deepest minima are reached at $q=0,\pm 4K$. It would be interesting to know the orientation of the angular momentum ${\bf I}$, in the potential minima. To achieve this goal, it is useful to write the boson expansion of the a.m. components in a different, otherwise equivalent, form:
\begin{eqnarray}
\hat{I}_{+}&=&\frac{i}{k'}\left[I(-d+ck^2)s-(c-d)\frac{d}{dq}\right],\nonumber\\
\hat{I}_{-}&=&\frac{i}{k'}\left[I(d+ck^2)s+(c+d)\frac{d}{dq}\right],\nonumber\\
\hat{I}_1&=&Icd-s\frac{d}{dq}.
\end{eqnarray}
From here it is obvious that for $q=\pm 2K$ (i.e. $\varphi =\pm\pi$) $I_1=-I$, while for $q=0$ (which means $\varphi=0$) $I_1=I$. 

Expanding the potential $V(q)$ up to the second order in $q$, one obtains the equation of a harmonic oscillator leading to the spectrum for $H_{rot}$:
\begin{equation}
E_{n}=A_1I^2-(2I+1)A_1j_1-IA_2j_2+\hbar\omega(n+1/2)+\sum_{i=1,2}A_i{j_i}^2,
\end{equation} 
where the new frequency has the expression:
\begin{eqnarray}
\omega&=&\left[\left((2I+1)(A_2-A_1-\frac{A_2j_2}{I})+2A_1j_1\right) \left((2I+1)(A_3-A_1)+2A_1j_1 \right) \right. \nonumber\\
&-&\left.(A_3-A_1)(A_2-A_1-\frac{A_2j_2}{I})\right]^{1/2}.
\end{eqnarray}
Following the same procedure as before, we may expand the potential around the local minimum, $q=2K$, and keep only up to the quadratic term, we obtain the following quantal energies:
\begin{equation}
E^{1}_{n}=-\frac{(2I+1)^2}{2}v+\sqrt{(1+v)(u+v)(2I+1)^2-u}\left(n+\frac{1}{2}\right),
\label{enequa}
\end{equation}
where the notation $v=2v_0/(2I+1)$ was used.

We recall now that the true Hamiltonian is $\hat{H}_{rot}$, related with $\hat {H}^{\prime}$ through Eq.(2.4). Thus, the final spectrum has the expression:
\begin{equation}
E^{\prime}_{n}=A_1I^2+(2I+1)A_1j_1-IA_2j_2+\hbar\omega^{\prime}(n+1/2)+\sum_{i=1,2}A_i{j_i}^2.
\end{equation} 
where the frequency $\omega^{\prime}$ is defined by:
\begin{eqnarray}
\omega^{\prime}&=&\left[\left((2I+1)(A_2-A_1-\frac{A_2j_2}{I})-2A_1j_1\right)
\left((2I+1)(A_3-A_1)-2A_1j_1 \right) \right. \nonumber\\
&-&\left.(A_3-A_1)(A_2-A_1-\frac{A_2j_2}{I})\right]^{1/2}.
\end{eqnarray}

\begin{figure}
\includegraphics[width=0.4\textwidth]{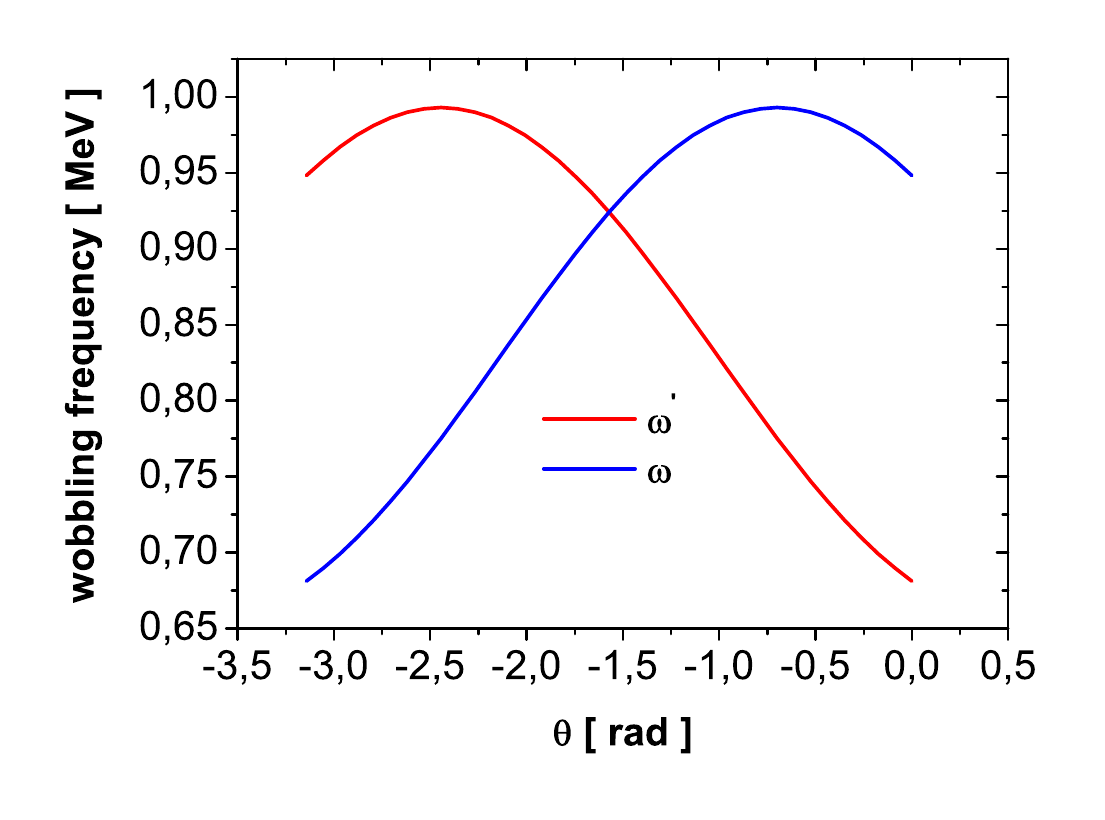}
\caption{(Color online) The phonon energies $\omega$ and $\omega'$ are plotted as function of $\theta$, defining the orientation of {\bf j} in the plane XOY, for the moments of inertia ($MoI$) 
: ${\cal J}_1:{\cal J}_2:{\cal J}_3=91:9:51\hbar^2 MeV^{-1}$ and j=13/2, I=55/2.}
\label{Fig.6}
\end{figure}
Of course, the phonon energies depend on the angle $\theta$ defining the components $j_1$, and $j_2$ of the single particle angular momentum. It is worth remarking that while the phonon energy $\omega$ defined inside the deepest well exhibits a maximum at about $-\pi/6$, the energy of the phonon defined in the local minimum shows a maximum at $-5\pi/6$ (see Fig. 6). The  two frequencies are equal for about $\theta = \pi/2$.
\renewcommand{\theequation}{5.\arabic{equation}}
\setcounter{equation}{0}
\section{Classical description}
The classical picture is obtained by the dequantization  procedure, which consists in replacing the operators $\hat{I}_k$ with the classical component of the angular momentum, $I_k$, and the algebra multiplication by:
\begin{equation}
\left[, \right]\to-i\left\{, \right\},
\end{equation}
with the notation $\{,\}$ for the Poisson bracket.
Let us now denote by $\varphi_k$ the conjugate coordinate of $I_k$.
Thus, the classical counterpart of the Hamiltonian $\hat{H}'$ is:

\begin{equation}
H'=I_2^2+uI_3^2+2v_0I_1,
\label{haspri}
\end{equation}
and one easily finds that:
\begin{equation}
\left\{I_k,H'\right\}=\stackrel{\bullet}{I_k},
\end{equation}
which leads to the following equations of motion:
\begin{eqnarray}
\stackrel{\bullet}{x}_1&=&2(1-u)x_2x_3,\nonumber\\
\stackrel{\bullet}{x}_2&=&2(x_1u-v_0)x_3,\nonumber\\
\stackrel{\bullet}{x}_3&=&-2(x_1-v_0)x_2,
\label{eqm}
\end{eqnarray}
where, for simplicity, the notation $x_k=I_k$, k=1,2,3 has been used. Also, the symbol $"\bullet"$ is used for the first derivative with respect to time. 
Using the equations of motion (\ref{eqm}), one proves that there are two constants of motion:
\begin{eqnarray}
E&=&x_2^2+ux_3^2+2v_0x_1,\nonumber\\
I^2&=&x_1^2+x_2^2+x_3^2.
\label{const}
\end{eqnarray}
This is a reflection of the fact that the energy, and the angular momentum are conserved.
The above equation allows us to express $x_2$, and $x_3$ in terms of $x_1$. Making the time derivative of the first equation (\ref{eqm}), and inserting the expressions of $x_2$, and $x_3$ in the resulting equation, one obtains the final equation for $x_1$:
\begin{equation}
\stackrel{\bullet\bullet}{x}_1+a_3x_1^3+a_2x_1^2+a_1x_1+a_0=0,
\label{eleq}
\end{equation}
where the coefficients have the expressions:
\begin{eqnarray}
a_3&=&8u,\nonumber\\
a_2&=&-12v_0(1+u),\nonumber\\
a_1&=&16v_0^2-8uI^2+4E(1+u),\nonumber\\
a_0&=&-8v_0E+4v_0(1+u)I^2.
\end{eqnarray}
We recognize in (\ref{eleq}) the differential equation for the elliptic functions of the first kind. Their explicit expressions can be obtained from the equations of motion (\ref{eqm}). Indeed, from (\ref{const}) one obtains:
\begin{eqnarray}
x_2&=&\left(1-u\right)^{-1/2}\left(ux_1^2-2v_0x_1+E-uI^2\right)^{1/2},\nonumber\\
x_3&=&\left(1-u\right)^{-1/2}\left(-x_1^2+2v_0x_1-E+I^2\right)^{1/2},
\end{eqnarray}
and then the first equation (\ref{eqm}) can be integrated with the result:
\begin{eqnarray}
t-t_0&=&\int_{x_{10}}^{x_1}\frac{dx}{2\sqrt{-u(x-\alpha_1)(x-\alpha_2)(x-\alpha_3)(x-\alpha_4)}}\nonumber\\
&\equiv &\frac{1}{\sqrt{C}}F(\varphi,k),
\label{integ}
\end{eqnarray}
where $\alpha_1, \alpha_2 $ are the roots of the equation $x_2=0$, while $\alpha_3, \alpha_4$, for $x_3=0$:
\begin{eqnarray}
\alpha_{1,2}&=&v_0\pm\left[v_0^2-u(E-I^2u)\right]^{1/2},\nonumber\\
\alpha_{3,4}&=&v_0\pm\left[v_0^2-E+I^2)\right]^{1/2}.
\end{eqnarray}
The limits for the integral (\ref{integ}) are chosen such that the integrand is a real number for any $x\in (x_{10},x_1]$. Obviously, the  integral (\ref{integ}) depends on the relative position of the poles $\alpha_i, i=1,2,3,4)$. The argument $\varphi$ involved in the elliptic function is defined as:
\begin{equation}
\varphi=\left\{\begin{matrix}\arcsin{k_1}, & if\rm{~ all~} \alpha_i \rm{~ are~ real}, \cr
                             \arctan{k_1},\; & \rm{if\; two\; }\alpha_i \rm {\;are\;complex\;numbers}\end{matrix} \right.
\label{fik1k2}
\end{equation}
The explicit expressions of $C,k_1^2, k^2$ are given in Ref.\cite{Rad98}.
Equation (\ref{integ}) can be reversed, and the result is a function $x_1(t)$, which is periodic, with the period:
\begin{equation}
T=\frac{\pi}{\sqrt{C}}{ _{2}F_{1}}\left(\frac{1}{2},\frac{1}{2},1;k^2\right).
\end{equation}
Here, the standard notation for the confluent hypergeometric function ${ _{2}F_{1}}\left(a,b,x;x\right)$ has been used.
In a similar way one may find the functions $x_2(t)$, and $x_3(t)$. The set of points $(x_1(t),x_2(t),x_3(t))|_{t}$ defines the classical trajectory which can further be quantized. Indeed, let $P_0$ be an extremal point on the sphere of the radius $I$, to which the energy $E_0$ corresponds. Let us now consider the trajectory to be quantized, characterized by the energy E and  surrounding $P_0$. 
 Consider the calotte bordered by the chosen trajectory, whose area defines the classical action. The quantization consists in restricting the action to be an integer multiple of $2\pi$.
\begin{equation}
{\cal L}(E)=\int \Omega =\int_{E_0}^{E}\int_{0}^{T}dE'dt'=\int_{E_0}^{E}T(E')dE'=2\pi n.
\end{equation}
From here, one easily finds:
\begin{equation}
\frac{\partial {\cal L}}{\partial E} =T(E)=\frac{\partial {\cal L}(E)}{\partial n}\frac{\partial n}{\partial E},\;\;\frac{\partial E}{\partial n }=\frac{2\pi}{T(E)}.
\end{equation}
It results that a linear dependence of $E$ on $"n"$ is obtained when $T(E)$ is approximated by its zero order expansion around $E_0$. In this case $E$ is given by the harmonic approximation.
For an even-even system the period expansion in terms of energy has been performed in Ref.\cite{Rad98}. 

Here we adopt a different procedure to obtain the harmonic motion of the even-odd system. 
Several situations are considered:
\begin{figure}[h!]
\includegraphics[width=0.4\textwidth]{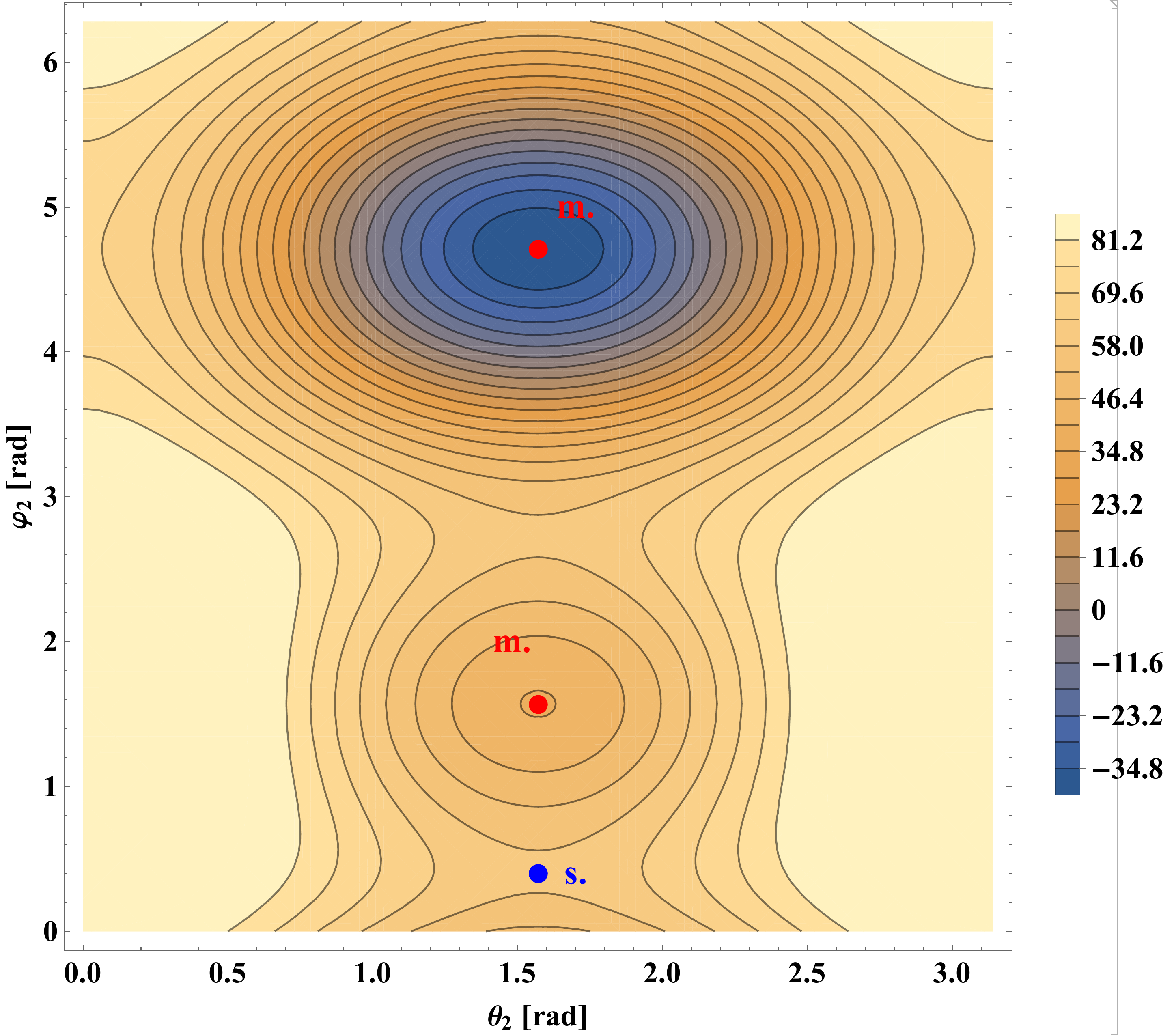}
\caption{(Color online) Contour plot of the case A) for I=19/2, j=11/2 and $\theta=70^0$ of $^{135}$Pr corresponding to the MoI's: ${\cal I}_1:{\cal I}_2:{\cal I}_3=10: 40 : 20 \hbar^2Mev^{-1}$. The minima (m.) and the saddle points(S.) are also mentioned}
\label{Fig.7}
\end{figure}

A1) Indeed, changing the Cartesian to the polar coordinates:
\begin{equation}
x_2=I\cos\theta_2,\;x_3=I\sin\theta_2\cos\varphi_2,\; x_1=I\sin\theta_2\sin\varphi_2,
\end{equation}
which is convenient in the case when the maximal MoI corresponds to the 2-axis, the energy function $H'$ can be expressed only in terms of the canonical conjugate coordinates $(x_2,\varphi_{2})$:
\begin{eqnarray}
H'&=&x_2^2\left(1-u\cos^2\varphi_2-\frac{v_0}{I}\sin\varphi_2\right)\nonumber\\
&+&uI^2\cos^2\varphi_2+2v_0I\sin\varphi_2.
\end{eqnarray}
The function $H'$ has a minimum in $(x_2,\varphi_2)=(0,-\frac{\pi}{2})$.
In the minimum point, the second derivatives of $H'$ have the values:
\begin{eqnarray}
\frac{\partial^2H'}{\partial x_2^2}\left|_{m}\right.&=&2\left(1+\frac{v_0}{I}\right),\nonumber\\
\frac{\partial^2H'}{\partial \varphi_{2}^{2}}\left|_{m}\right.&=&2\left(u+\frac{v_0}{I}\right)I^2.
\end{eqnarray}
Also the minimal value of $H'$ is:
\begin{equation}
H'\left|_{m}=-2v_0I\right..
\end{equation}
Denoting by $(\bar{x}_2,\bar{\varphi}_2)$ the deviation of the current coordinates form those of the minimum point, the second order expansion of $H'$ looks like:
\begin{equation}
H'=-2v_0I+\left(1+\frac{v_0}{I}\right)\bar{x}_2^2+\left(u+\frac{v_0}{I}\right)I^2\bar{\varphi}_2^2.
\end{equation}
This describes a harmonic oscillator of a frequency:
\begin{equation}
\omega=2\sqrt{(1+v)(u+v)I^2},
\end{equation}
with $ v=\frac{v_0}{I}$. By quantization, the spectrum corresponding to $ H'$ coincides with that from (\ref{enequa}), provided the approximation $I+\frac{1}{2}\approx I$ is adopted. 

In the minimum point, the angular momentum components are:
\begin{equation}
(x_1,x_2,x_3)=(-I,0,0)_{m},
\end{equation}
while the energy is: $E_m=-2vI^2$.

A2) One may check that $(0,\frac{\pi}{2})$ is also a minimum of $H'$, in which the angular momentum is $(I,0,0)$, while the energy has the expression $E_m=2vI^2$.
The second order expansion of $H'$ is :
\begin{equation}
H'=2v_0I+(1-v)\bar{x}_2^2+I^2(u-v)\bar{\varphi}_2^2.
\end{equation}
This describes an harmonic oscillation of frequency:
\begin{equation}
\omega=2\sqrt{(1-v)(u-v)I^2}.
\end{equation}
This frequency coincide with the quantal frequency $\omega^{\prime}$  given by Eq.(4.5), if we adopt the approximation $I+\frac{1}{2}\approx I$, which is valid for a large $I$.

A3) Another pair of conjugate, and stationary variables, which might be minimum for the energy function is:
\begin{equation}
(x_2,\varphi_2)_s=(0,\arcsin\left(\frac{v_0}{Iu}\right)).
\end{equation}
To this, it corresponds  the harmonic Hamiltonian;
\begin{equation}
H^{\prime}_{h}=uI^2+\frac{v_0^2}{u}+(1-u)\bar{x}_2^2-u\left(I^2-\frac{v_0^2}{u^2}\right)\bar{\varphi}_2^2.
\end{equation}
If $0<u<1$, the mentioned stationary point is a saddle point, to which the following angular momentum  corresponds: $(x_1,x_2,x_3)=(\frac{v_0}{u}, 0, \sqrt{(I^2-\frac{v_0^2}{u^2}})_{s}$,
with the energy of: $E_s=(u+\frac{v^2}{u})I^2$. In the case $-1<u<0$ the above mentioned stationary point is a minimum.  The trajectories corresponding to the situations labeled by A1)-A3) are visualized by the contour plot given in Fig. 7. One notices that the trajectories of energies close to a minimum, surround exclusively that minimum, while  due to the tunneling effect, the trajectories of large energies surround all minima.

B1) Choosing now the 3-axis as quantization axis, and the corresponding polar coordinates:
\begin{equation}
x_1=I\sin\theta_3\cos\varphi_3,\;x_2=I\sin\theta_3\sin\varphi_3,\;x_3=I\cos\theta_3,
\end{equation}
the Hamiltonian $H'$ can be expressed in terms of the canonical conjugate variables $(x_3,\varphi_3)$:
\begin{equation}
H'=x_3^2\left(u-\sin^2\varphi_3-\frac{v_0}{I}\cos\varphi_3\right)+I^2\sin^2\varphi_3+2v_0I\cos\varphi_3.
\end{equation}
One stationary point which might be a minimum is $(x_3,\varphi_3)=(0,\pi)$. The corresponding angular momentum components are
$(x_1,x_2,x_3)=(-I,0,0)_m$. Therefore, the angular momentum is oriented along the 1-axis. The minimum energy is $E_{m}=-2v_0I$.
The harmonic Hamiltonian, i.e. the second order expansion of $H'$ around $(0,\pi)$, is:
\begin{equation}
H_h^{\prime}=-2v_0I+\left(u+\frac{v_0}{I}\right)\bar{x}_3^2+I^2\left(1+\frac{v_0}{I}\right)\bar{\varphi}_3^2.
\end{equation}
Although the harmonic Hamiltonian is different from that from the case A1), the two scenarios provide the same angular frequencies, but the canonical conjugate variables are interchanged.
\begin{figure}[h!]
\includegraphics[width=0.4\textwidth]{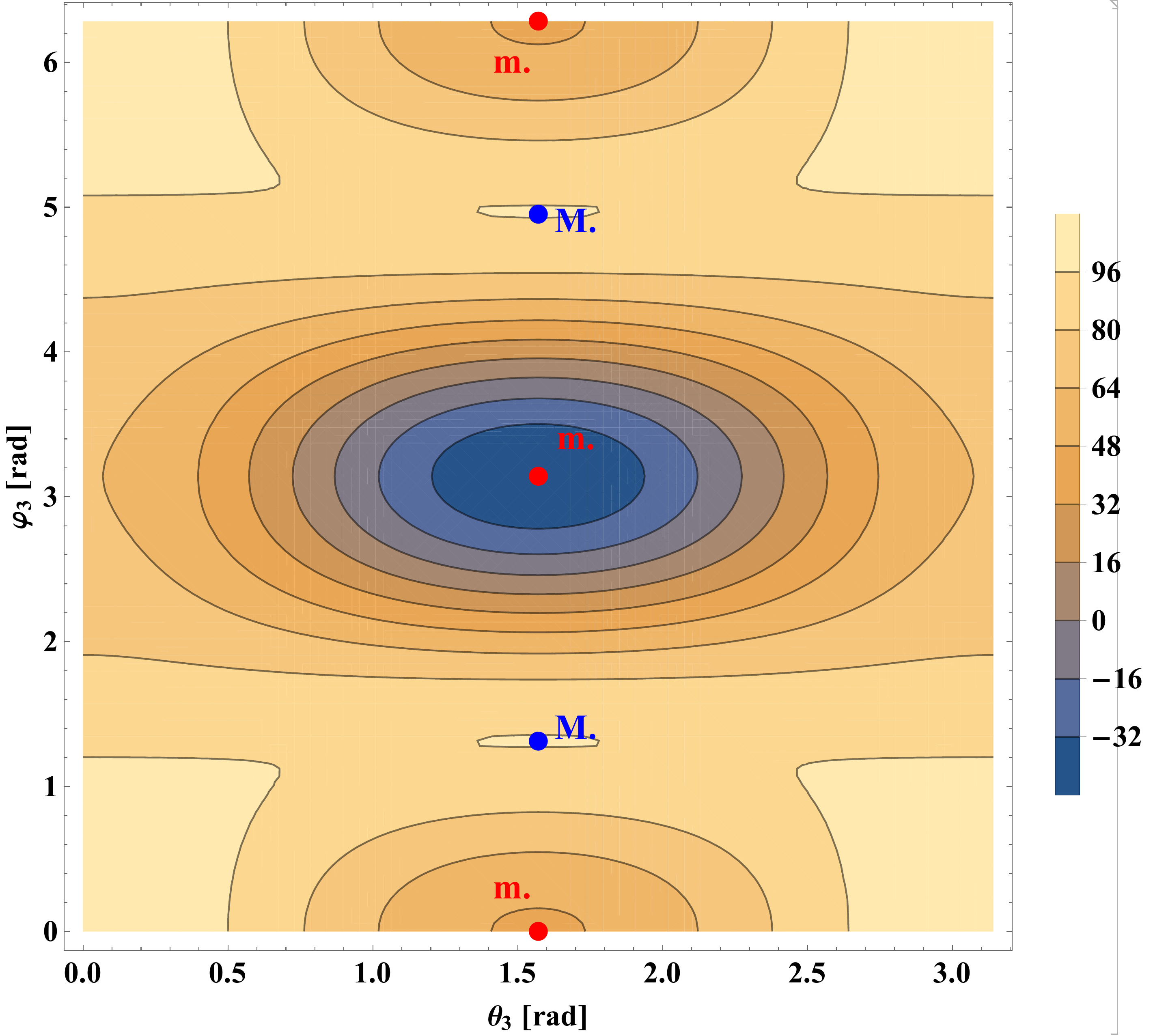}
\caption{(Color online)Contour plot of the case B) for I=19/2, j=11/2 and $\theta=70^0$ of $^{135}$Pr corresponding to the MoI's: ${\cal I}_1:{\cal I}_2:{\cal I}_3=10: 20: 40 \hbar^2Mev^{-1}$. The minima (m.) and maxima (M.) are also mentioned.}
\label{Fig.8}
\end{figure}

B2) Similarly, one shows that $(x_3,\varphi_3)=(0,0)$ is a minimum, which results the quadratic expansion: 
\begin{equation}
H_h^{\prime}=2v_0I+\left(u-\frac{v_0}{I}\right)\bar{x}_3^2+I^2\left(1-\frac{v_0}{I}\right)\bar{\varphi}_3^2,
\end{equation}
with the stationary angular momentum ($(x_1,x_2,x_3)=(I,0,0)_m$, and the corresponding energy $E_m=2v_0I$.
The harmonic frequency determined by $H_h^{\prime}$:
\begin{equation}
\omega=2\sqrt{(1-v)(u-v)I^2}.
\end{equation}

B3) In this case, the stationary point is  $(x_3,\varphi_3)=(0,\arccos\frac{v_0}{I})$, which leads to $(x_1,x_2,x_3)=(v_0,\sqrt{I^2-v_0^2},0)_M$.
The corresponding  quadratic expansion of $H'$ is:
\begin{equation}
H^{\prime}_{h}=I^2+v_0^2+(u-1)\bar{x}_3^2+(v_0^2-I^2)\bar{\varphi}_3^2,
\end{equation}
which indicates that the stationary point is a maximum point with the critical energy equal to $E_M=I^2(1+v^2)$. 
The trajectories determined by the circumstances specified by the cases B1)-B3) are represented in the contour plot from Fig. 8.

C1) If the maximal MoI corresponds to the 1-axis, then we 
choose this as quantization axis, and the polar coordinates:
\begin{equation}
x_1=I\cos\theta_1,\;x_2=I\sin\theta_1\cos\varphi_1,\;x_3=I\sin\theta_1\sin\varphi_1.
\end{equation}
The Hamiltonian becomes:
\begin{equation}
H^{\prime}=\left(\cos^2\varphi+u\sin^2\varphi\right)\left(I^2-x_1^2\right)+2v_0x_1.
\end{equation}
This has a stationary point in $(x_1,\varphi_1)=\left(\frac{v_0}{u},\frac{\pi}{2}\right)$.
For $0<u<1$, this is a saddle point for $H'$, as suggested by the second order expansion:
\begin{equation}
H^{\prime}_{h}=uI+\frac{v_0}{u}-u\bar{x}_1^2+(1-u)\left(I^2-\frac{v_0^2}{u^2}\right)\bar{\varphi}_1.
\end{equation}
In the case $-1<u<0$ the mentioned stationary point is a minimum.
The corresponding angular momentum and energy are: $(x_1, x_2, x_3)=(\frac{v_0}{u}, 0, \sqrt{(I^2-\frac{v_0^2}{u^2}})_{s}$, and $E_{s}=(u+\frac{v^2}{u})I^2.$

\begin{figure}[h!]
\includegraphics[width=0.4\textwidth]{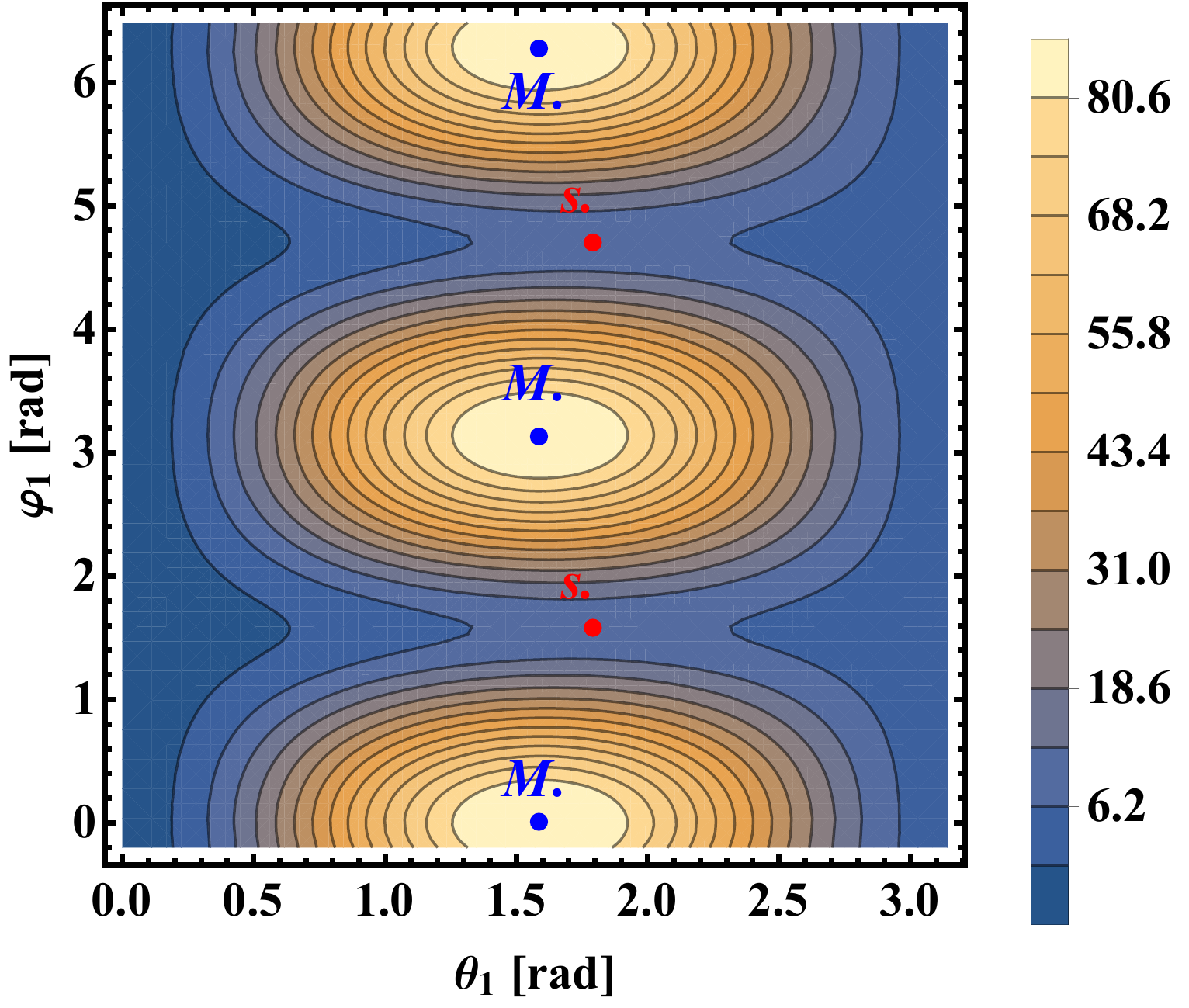}
\caption{(Color online) The contour plot of the case C) for I=19/2 of $^{135}$Pr corresponding to the MoI's, and $\theta$ determined by the adopted fitting procedure. The maxima and saddle points are also mentioned.}
\label{Fig.9}
\end{figure}

C2) The stationary point $(v_0,0)$  is a maximum, with the angular momentum $(x_1, x_2, x_3)=(v_0, \sqrt{I^2-v_0^2},0)_{M}$, and energy $E_M=I^2+v_0^2$. The quadratic expansion around this point is:
\begin{equation}
H^{\prime}_{h}=I^2+v_0^2 -\bar{x}_1^2+(u-1)(I^2-v_0^2)\bar{\varphi}_1^2.
\end{equation}
Pictorially, this case is shown in Fig.9. The maxima points indicate that rotations around the cranking axis are forbidden.
Concluding this analysis, there are six  minima, the cases A1), A2), B1), B2), and A3), C1 for $-1<u<0$, one maximum, the cases B3),C2), and one saddle point, the situations A3) and C1) for
$0<u<1$. The frequencies corresponding to the six minima are grouped in two pairs of degenerate frequencies, and moreover the  frequency showing up in the cases A1), and B1) is equal to the one provided by the quantal description for the deepest minimum of the potential energy. The other two degenerate minima, A2) and B2), produce a frequency equal to the one showing up in the quantal description for the local minimum. In the minimum points, the total angular momentum is oriented along a principal axis,namely the 1-axis, while for the maximum and the saddle point is located in a principal plane. For $-1<u<0$ the minima characterizing the situations A3) and C1) correspond to an angular momentum located in the plane $(x_1,x_3)$. It is worth mentioning that in the maximum point, the angular momentum is oriented along the 2-axis to which the maximal MoI corresponds. Therefore, the transverse wobbling is unstable.

\begin{figure}[h!]
\includegraphics[width=0.4\textwidth]{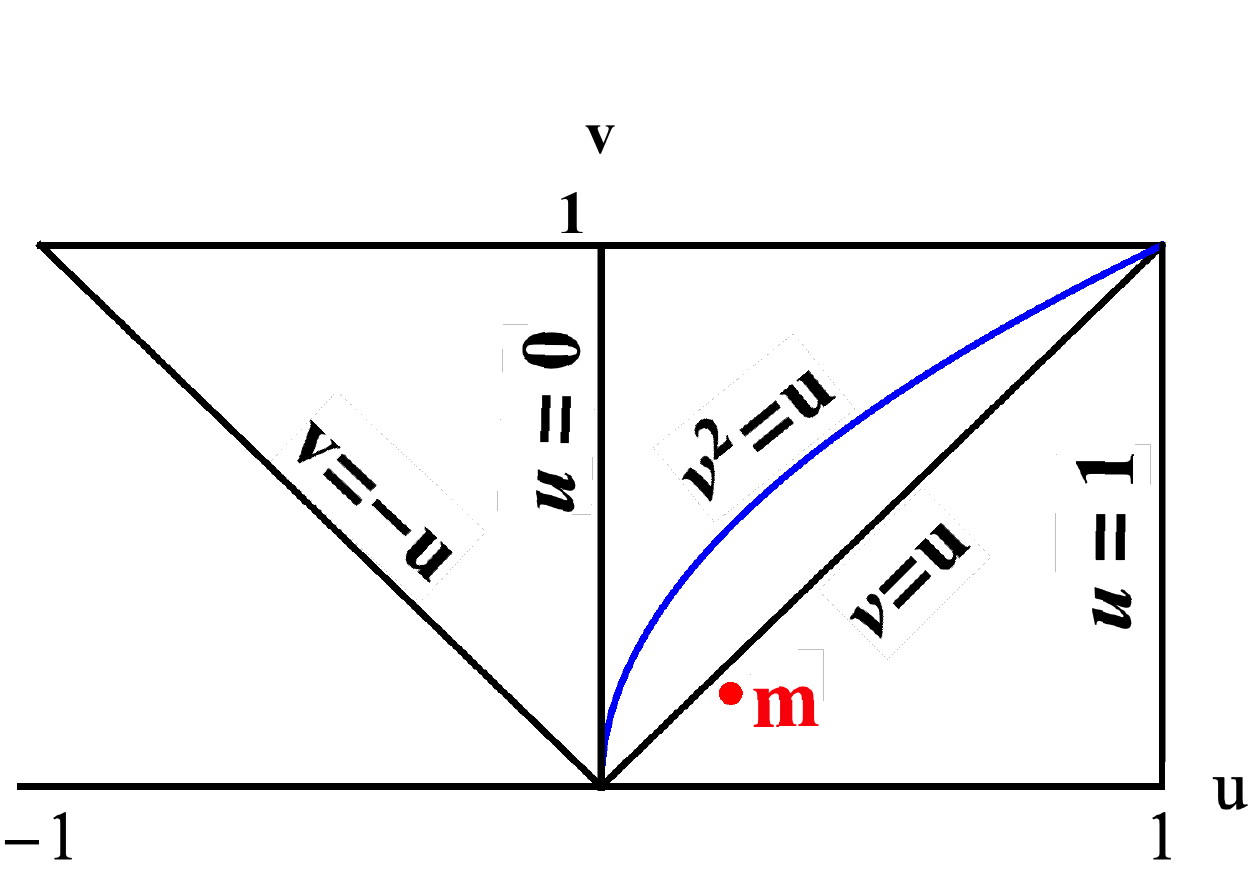}
\caption{(Color online) The phase diagram for I=15/2 of $^{135}$Pr. Using the MoI's, and $\theta$ determined by the adopted fitting procedure, we calculated the coordinates of the minimum point, the result being shown  by a red and full circle having a lowercase $m$.}
\label{Fig.10}
\end{figure}
According to the contour plot of Fig. 9, all trajectories, determined by the conditions C1,C2), are meta-stable for $0<u<1$.
\subsection{The phase diagram}
The character of the stationary point to be minimum, maximum or saddle point is decided by the signs of the diagonal matrix elements of the Hessian: a) if all diagonal elements are positive, the stationary point is minimum; b) if all diagonal m.e. are negative, then we deal with a maximum, while c) it is a saddle point if one m.e. is positive and the other is negative.
Equating the Hessian to zero, one obtains the parameters $u$, and $v$ for which the critical points are degenerate. The resulting equations may be unified in a single formula:
\begin{equation}
(1-u)(1-v^2)(v^2-u^2)(v^2-u)=0.
\label{separ}
\end{equation} 
The last factor in the above equation is obtained by equating the critical energies $E_{M}$, and $E_{s}$. Each factor generates a curve, called separatrice, in the parameter space spanned by (u,v).
As shown in Fig. 10, the separatrices are bordering  manifolds defining  unique nuclear phases characterized by a specific portrait of the stationary points. Indeed, among the factors involved in Eq.(\ref{separ}), we recognize those defining the two wobbling frequencies. On the other hand a vanishing energy defines a Goldstone mode \cite{Gold} which, as a matter of fact, render evidently a phase transition.

\setcounter{equation}{0}
\renewcommand{\theequation}{6.\arabic{equation}}
\section{Electromagnetic transitions}
We are interested in describing the experimental data for the electric quadrupole intra- and inter-band transitions as well as the magnetic dipole transitions. We begin with the electric transitions. Aiming at this goal, we need the wave functions describing the involved states, and the transition quadrupole operator. The wave function for an I-state is the solution of the 
Schr\"{o}dinger equation for the given total angular momentum, I. Note that the wave function is degenerate with respect to "M", the projection of I on the x-axis, in the laboratory frame. Since the ground state is the vacuum state for the wobbling phonon operator, and moreover, in the minimum point of the constant energy surface, the a.m. projection on the one-axis of the rotating frame is equal to -I, it results that the K quantum number is equal to -I. Therefore the solution of the Schr\"{o}dinger equation must be labeled by  the mentioned  quantum numbers, i.e. 
\begin{equation}
\Psi_{IM}=\Phi_{I,-I}|IM,-I\rangle, \;\;\rm{with}\;\;
|IMK\rangle =\sqrt{\frac{2I+1}{8\pi^2}}D^I_{M,K}.
\end{equation}
Here we consider the first two wobbling bands as signature partner bands, the arguments being in detail given in Ref.\cite{Rad201}. More specifically, the spin sequence of the first band is 
$j+R$, for R=0,2,4,... , while for the second band the spin succession is $j+R$, with R=1,3,5,....  Note that the quadrupole inter-band transition is forbidden since, for the states mentioned above, we have $\Delta K=1$. In this case, considering the component $K=-I+1$ in one of the involved states, i.e. $\Phi_{I,-I}$, is necessary. The quadrupole transition operator is taken as:
\begin{equation}
{\cal M}(E2;\mu)=\sqrt{\frac{5}{16\pi}} e\left(Q_{20}D^2_{\mu 0}+Q_{22}D^2_{\mu 2}+Q_{2-2}D^2_{\mu -2}\right),
\end{equation}
where $Q_{20}$, and $Q_{2\pm 2}$, denote the $K=0$, and $K=\pm 2$ components of the quadrupole transition operator, respectively.  Note that since the intrinsic component of the wave function depends on one of the conjugate variables $q$ and $d/dq$, that is $q$, we must express the quadrupole operators in terms of the q variable. This will be achieved by writing Q-s in the space of angular momentum and then use the Bargmann representation of the a.m. components. Thus we have:
\begin{equation}
Q_0=\left(-\frac{1}{4}\sqrt{\frac{2}{3}}\left(\hat{I}_{+}\hat{I}_{-}+\hat{I}_-\hat{I}_+\right)+\sqrt{\frac{2}{3}}\hat{I}_1^2\right)\bar{Q}_0,\;\;
Q_{\pm 2}=\frac{1}{2}\hat{I}_{\pm}^{2}\bar{Q}_2.
\label{qtran}
\end{equation}
The quadrupole operator was previously written in terms of angular momentum in Ref.\cite{Davyd}, where the quadrupole-quadrupole coupling of a particle and a triaxial rotor-core is studied.
Recently, the static quadrupole moment in a wobbling state was considered in the space of intrinsic  angular momenta \cite{Chen}.

The expressions of the intrinsic quadrupole operators in terms of the conjugate variables $q$ and $d/dq$ are given in Appendix A.

 Here we propose an alternative method to calculate the necessary matrix elements, which is easier to handle. We use the expression (\ref{qtran}), and evaluate the result of acting on the system wave function with the spherical components of ${\bf I}$. The nice feature of this procedure is that any reduced m.e. has  the overlap of the initial and final intrinsic states as a common factor. As an example, the overlap factors for inter-band and intra-band transitions, are:
\begin{equation}
\langle \Phi_{I,-I}|\Phi_{I-1,-I+1}\rangle =\frac{2\sqrt{\bar{\omega}_{I}\bar{\omega}_{I-1}}}{\bar{\omega}_{I}+\bar{\omega}_{I-1}},\;\;
\langle \Phi_{I,-I}|\Phi_{I-2,-I+2}\rangle =\frac{2\sqrt{\bar{\omega}_{I}\bar{\omega}_{I-2}}}{\bar{\omega}_{I}+\bar{\omega}_{I-2}}.
\end{equation}
We checked numerically the fact that the overlap factors are very close to unity. For this reason we approximate them to one.
Thus, the  matrix elements involved in the equation defining the reduced transition probabilities are analytically expressed. Note that acting with the operator $Q_{2\mu}$ on the intrinsic wave function the a.m. is preserved but the $K$ quantum number is changed by $\mu$ units. The change of a.m. is due to the overlap factor which modifies also the K quantum number by 2 units. Due to this feature 
the variation $\Delta K$ for initial and final states might exceed $\mu$. However, we should keep in mind that the tensor properties of the operator $Q_{2\mu}$ specific to the laboratory frame, are lost when one passes to the rotating frame. On the other hand, the intrinsic wave function does not have $K=-I$, this being fulfilled for the classical minimal energy but not within the quantal picture. However, the conservation rules as well as the Wigner-Eckart theorem hold due to the laboratory frame wave factor.

The matrix elements for the intra-band transitions are:
\begin{eqnarray}
&&\langle \Psi_{I}||{\cal M}(E2)||\Psi_{I-2}\rangle =\sqrt{\frac{5}{16\pi}}e\left\{\bar{Q}_2\frac{1}{2}\left[C^{I \;\;2\; I-2}_{I \;-2\; I-2}
\left(\sqrt{6(I-1)(2I-3)}+\frac{2I^2-2I+5}{\sqrt{(I-2)(2I-5)}}\right)\right.\right.\nonumber\\
&+&\left.\left.C^{I \;\;2\; I-2}_{I-4 2 I-2}\left(\sqrt{I(2I-1)}+\frac{2I^2-2I+5}{\sqrt{I(2I-1)}}\right)\right]\right.
+\left.\bar{Q}_0C^{I \;\;2 \;I-2}_{I-2 \;0\; I-2}\frac{1}{\sqrt{6}}(2I^2-5I+5)\right\},
\end{eqnarray}
while those determining the inter-band transitions have the expressions \footnote{$^{*}$The convention of Rose \cite{Rose} for the reduced matrix elements has been used}:
\begin{eqnarray}
&&\langle \Psi_{I}||{\cal M}(E2)||\Psi_{I-1}\rangle =\sqrt{\frac{5}{16\pi}}e\left\{\bar{Q}_2\frac{1}{2}\left[C^{I \;\;\;2\; I-1}_{I-3 \; 2\;  I-1}
\left(\sqrt{I(2I-1)}+\frac{2I^2-3I+\frac{3}{2}}{\sqrt{I(2I-1)}}\right)\right.\right.\nonumber\\
&+&\left.\left.C^{I \;\;\;2\; I-1}_{I \; -2\;  I-2}\left(\sqrt{(I-1)(2I-3)}+\frac{2I^2-3I+\frac{3}{2}}{\sqrt{(I-1)(2I-3)}}\right)\right]\right.
+\left.\bar{Q}_0C^{I\;\;\;\; 2\;\;\; I-1}_{I-1 \; 0 \; I-1}\frac{1}{\sqrt{24}}\left(4I^2-6I+3\right)\right\}.
\end{eqnarray}
Furthermore, the reduced transition probabilities are readily obtained:
\begin{equation}
B(E2;I\to I')=\left[\langle \Psi_{I}||{\cal M}(E2)||\Psi_{I'}\rangle\right]^2.
\end{equation}
The factors $\bar{Q}_0$, and $\bar{Q}_2$ have the units of $\rm{e}.\rm{fm}^2/\hbar^2$, and are taken as free parameters.

The magnetic dipole transition operator is:
\begin{equation}
{\cal M}(M1;\mu)=\sqrt{\frac{3}{4\pi}}\mu_N\sum_{\nu}\left(g_R\hat{R}_{\nu}+g_j\hat{j}_{\nu}\right)D^{1}_{\mu \nu}
                \equiv M^{coll}_{1\mu}+M^{sp}_{1\mu},
\end{equation}
where $R_{\nu}$, and $j_{\nu}$ are the spherical components of the core and the odd nucleon angular momenta, respectively. $g_R$ and $g_j$ stand for the gyromagnetic factors of the core, and the coupled odd nucleon, respectively. Also, the standard notation for the Wigner function, $D^{J}_{MK}$, and for the nuclear magneton, $\mu_{N}$, are used.
To calculate the collective part of the transition matrix element, we need to express the wave function describing the odd system as a Kronecker product of the core, and the odd particle wave functions:
\begin{equation}
|IMK\rangle =\frac{1}{2j+1}\sum_{M_R,\Omega,R} C^{R\;j\;I}_{M_R\;\Omega\;M }|RM_RK\rangle \psi_{j\Omega}.
\end{equation}
By a direct manipulation, one finds: 
\begin{eqnarray}
\langle I||M^{coll}_{1}||I-1\rangle &=& \sqrt{\frac{3}{4\pi}}g_R\mu_{N}\frac{1}{2j+1}C^{I+j-1 1 I+j-1}_{I+j-2 1 I+j-1}
\left[(2I-1)(2J+2j-1)(I+j-1)(I+j)\right]^{1/2}\nonumber\\
&\times&W(I-1,j,1,I+j;I+j-1,I),
\end{eqnarray}
where the notation $W(a,b,c,d;e,f)$ stands for the Racah coefficient.

To calculate the reduced m.e. of the single particle M1 operator we need the wave function describing the odd proton whose a.m. is placed in the plane XOY making the angle $\theta$ with the axis OX. This function is obtained by rotating around the axis 3, the function $\psi_{j,j}$  associated with the odd proton having the a.m. along the 1-axis.
\begin{equation}
\psi_{j}^{\prime}=R_3(\theta)\psi_{jj}.
\end{equation}
The reduced m.e. of the single particle transition operator is:
\begin{eqnarray}
&&\langle I||M^{sp}_{1}||I-1\rangle =\sqrt{\frac{3}{4\pi}}g_j\mu_{N}C^{I\;\;1\;\;I-1}_{I\;-1\;-I}
\langle\psi_{jj}|R^{\dagger}_{3}(\theta)j_{-1}R_{3}(\theta)|\psi_{jj}\rangle\nonumber\\
&=&\frac{1}{\sqrt{2}}\langle \psi_{jj}|-\hat{j}_{1}\sin\theta+\hat{j}_{+}\frac{\cos\theta-1}{2}+\hat{j}_{-}\frac{\cos\theta+1}{2}|\psi_{jj}\rangle
=-j\sin\theta\sqrt{\frac{3}{8\pi}}\mu_{N}g_jC^{I\;\;1\;\;I-1}_{I\;-1\;-I}.
\end{eqnarray}
The gyromagnetic factors have the expressions:
\begin{equation}
g_{R}=\frac{Z}{A},\;
g_j=g_l+\frac{\frac{3}{4}+j(j+1)-l(l+1)}{j(j+1)}\frac{g_s-g_l}{2},
\end{equation}
where $g_l$, and $g_s$ stand for the orbital and spin free gyromagnetic factors, respectively.
Finally, the magnetic dipole reduced transition probability is given by:
\begin{equation}
B(M1;I\to I')=\left[\langle \Psi_{I}||{\cal M}(M1)||\Psi_{I'}\rangle\right]^2.
\end{equation}

\setcounter{equation}{0}
\renewcommand{\theequation}{7.\arabic{equation}}
\section{Results}
The formalism described in the previous sections was applied to $^{135}$Pr. The excitation energies in three bands, conventionally called band 1 (B1), band 2 (B2), and band 3 (B3), and the  electromagnetic properties of the states have been described by a simple Hamiltonian (2.1), associated with the even-even core and the odd proton, which stays in the orbital $h_{11/2}$.The core properties are
simulated by a triaxial core with the moments of inertia ${\cal J}_k$ (k=1,2,3), considered to be free parameters,  while the odd proton is rigidly coupled to the core, 
placed in the inertial plane (1,2) and having the polar angle $\theta$. Thus, the approach involves four free parameters ${\cal J}_k$  (k=1,2,3), and $\theta$, which were fixed by a least mean square procedure, fitting the excitation energies for the three wobbling bands in $^{135}$Pr.
{\scriptsize
\begin{table}[h!]
\begin{tabular}{|c|c|c|c|c|c|}
\hline
 ${\cal I}_1$& ${\cal I}_2$ &${\cal I}_3$& $\theta$ &nr. of&r.m.s.\\
$[\hbar^2/MeV]$           & $[\hbar^2/MeV]$&$[\hbar^2/MeV]$&  [degrees] & states& [MeV]              \\
\hline
91                        &   9                      &      51                   & -119 &20       &0.174\\
\hline
\end{tabular}
\caption{The MoI's, and the parameter $\theta$ as provided by the adopted fitting procedure. }
\label{Table 1}
\end{table}
}

\subsection{Energies}
In Ref.[28], the first two bands -TSD1 and TSD2- in the even-odd isotopes $^{161,163,165,167}$ Lu were interpreted as signature partner bands. The argument was that the potential well  is very deep and therefore comprises the energy levels of both bands, which results in having similar properties for the mentioned bands. Here, the potential well is also very deep and consequently we may consider the bands 1 and 2 as  partner bands with the signatures $+\frac{1}{2}$(favored) and $-\frac{1}{2}$ (unfavored),respectively. Moreover, the band 3 is interpreted as being an one phonon excitation of the band 2.

Thus, the excitation energies for the first three bands are obtained from Eq.(4.2):
\begin{eqnarray}
E^{exc;1}_{I} &=& A_1 I^ 2 +(2I+1)A_1 j_{1}-IA_2 j_2 + \omega_{I}/2-E_{11/2},\nonumber\\
\;\; I&=&R+j,\;\;R=0,2,4,...,\nonumber\\
E^{exc;2}_{I} &=& A_1 I^ 2 +(2I+1)A_1 j_{1}-IA_2 j_2 + \omega_{I}/2-E_{11/2},\nonumber\\
   I&=&R+j,\;\;R=1,3,5,...,\nonumber\\
E^{exc;3}_{I+1} &=& A_1 I^ 2 +(2I+1)A_1 j_{1}-IA_2 j_2 + 3\omega_{I}/2-E_{11/2},\nonumber\\
                 I&=&R+j,\;\;R=1,3,5,....
\label{exen}
\end{eqnarray}
As we already mentioned, the involved parameters were fixed by fitting the experimental excitation energies with those described by the above equations.
The parameters yielded by the fitting procedure are listed in Table I. With the parameters thus determined and Eq.(7.1), the excitation energies are readily obtained. They are visualized in Figs.11, 12, 13 and compared with the corresponding experimental data taken from Refs.\cite{Matta,Sen}. Note that in the mentioned figures the three bands are conventionally called as the first, the second and the third band, respectively. Results are compared with the experimental data \cite{Sen} for the bands yrast, one phonon wobbling (TW1) and the two phonon wobbling bands (TW2), respectively.
Our denomination is different from that used for the experimental bands, since in our case the second band is the signature partner band of the yrast band, while the third band is one phonon band built up on the base of the second band. From there, one sees the quality of the agreement with the data, that might be appraised by the r.m.s. of the deviation which is also given in Table I. This is slightly larger but comparable with the r.m.s.'s obtained in Refs.\cite{Chen1} ($\approx$0.160MeV) and \cite{Buda} ($\approx$0.150 MeV) by different methods. We may conclude that the agreement between theoretical, and experimental results is good.
\begin{figure}[h!]
\includegraphics[width=0.4\textwidth]{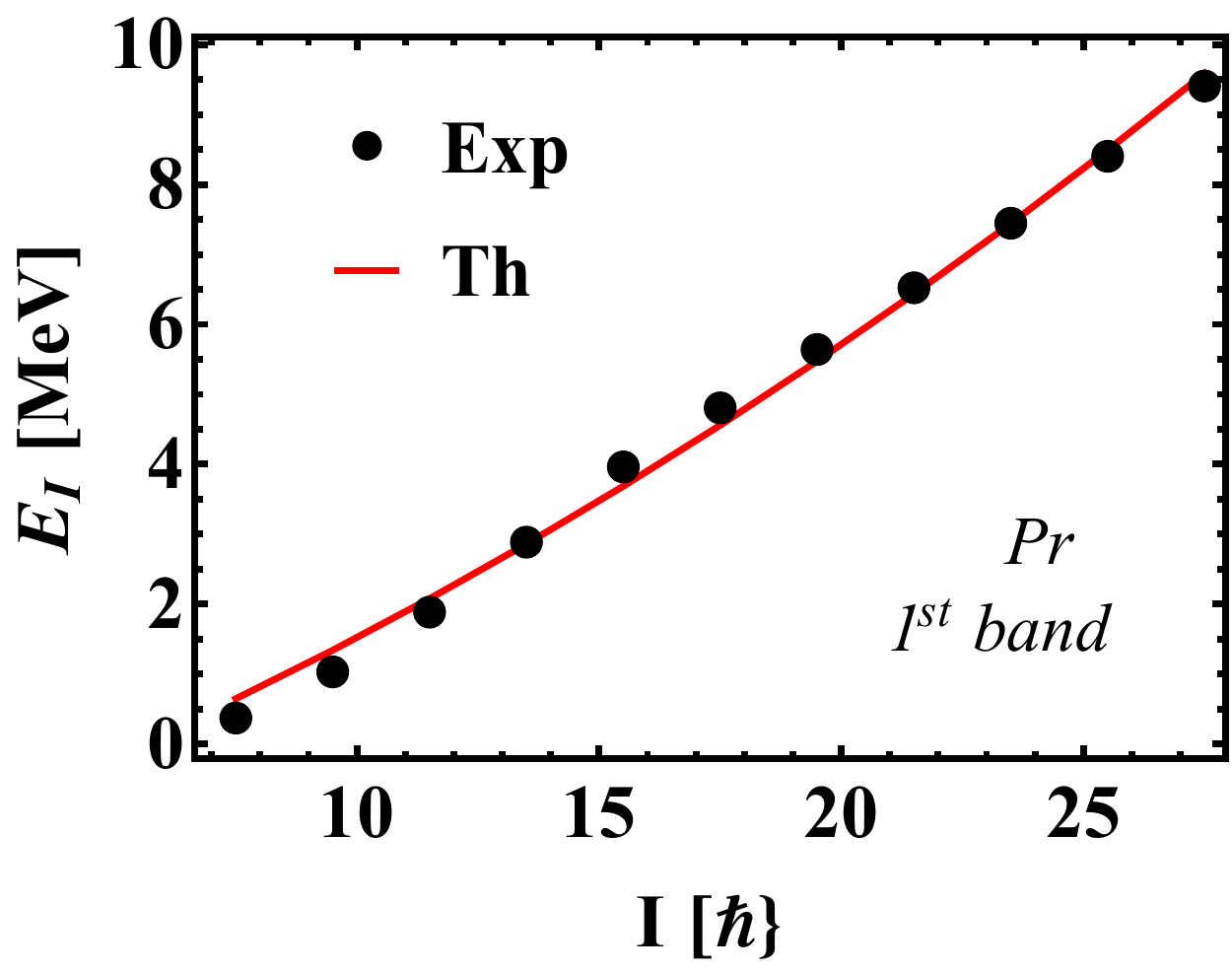}
\caption{(Color online) The excitation energies yielded by our calculations for the first band of $^{135}$Pr, using the MoI's, and $\theta$ determined by a fitting procedure,
are compared with the experimental excitation energies from the yrast band \cite{Sen}.}
\label{Fig.11}
\end{figure}

\begin{figure}[h!]
\includegraphics[width=0.4\textwidth]{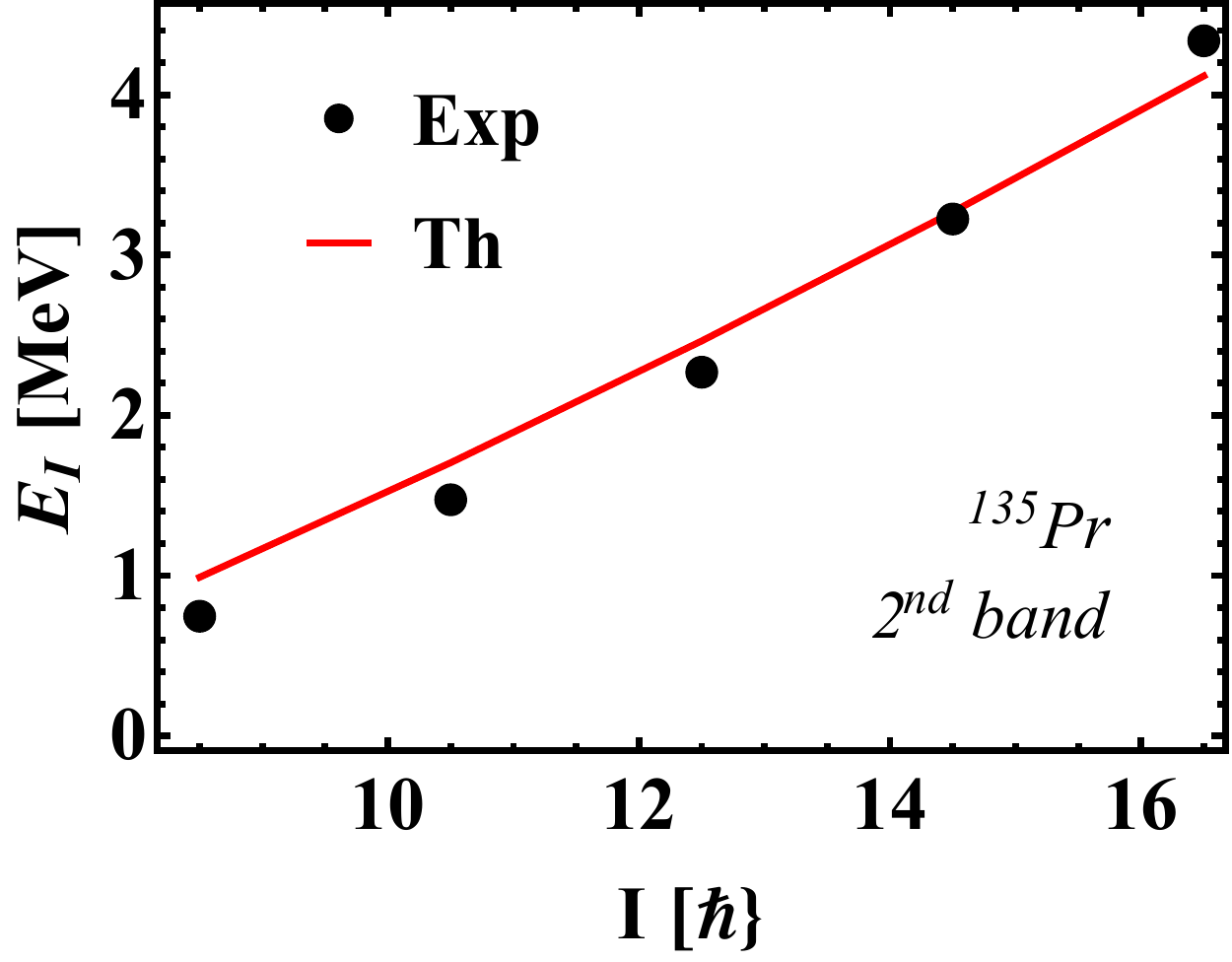}
\caption{(Color online) The excitation energies for the second band  of $^{135}$Pr, with the parameters determined as explained in the text, are compared with the experimental excitation energies from the TW1 band \cite{Sen}.
}
\label{Fig.12}
\end{figure}

\begin{figure}[h!]
\includegraphics[width=0.4\textwidth]{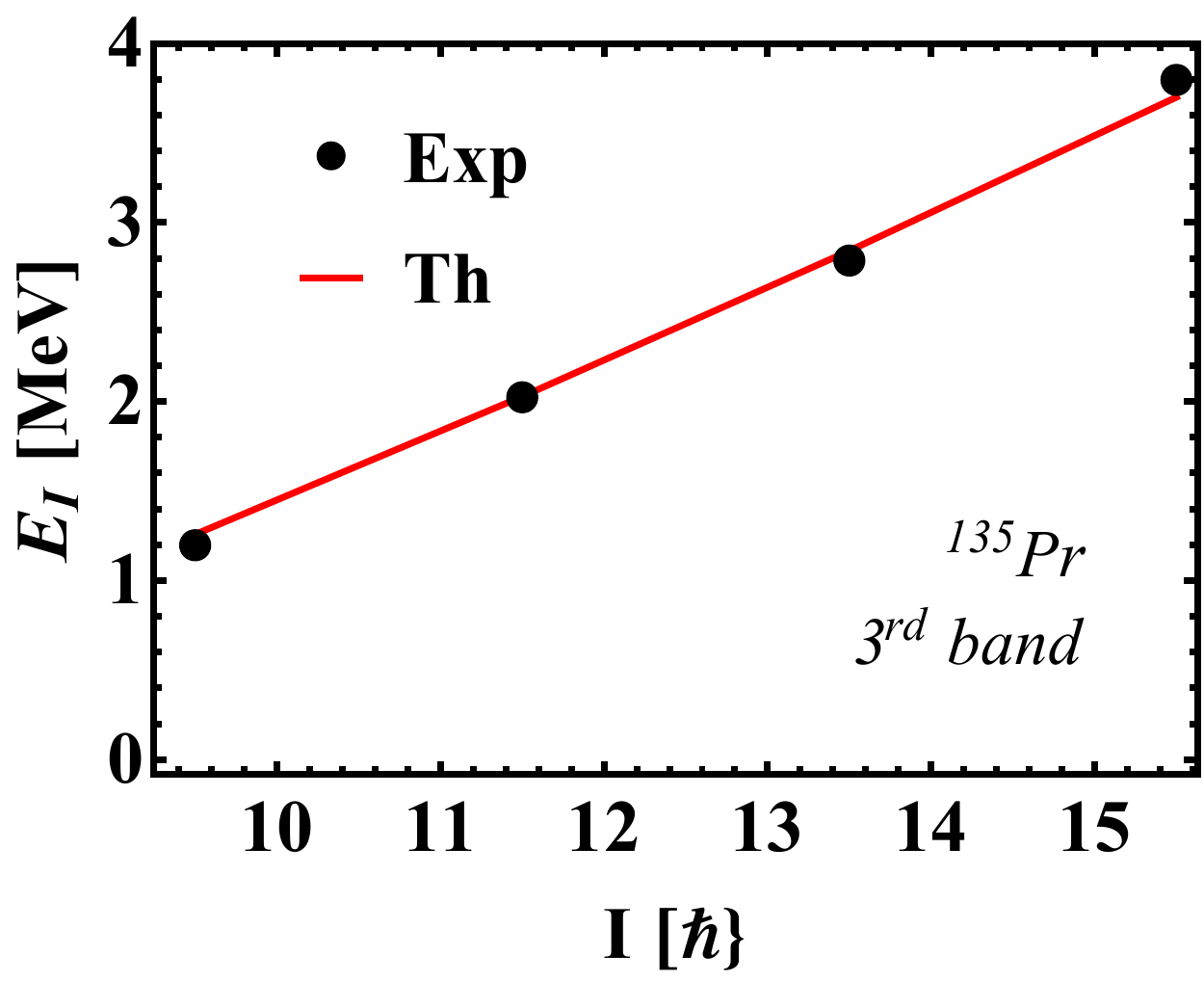}
\caption{(Color online) The excitation energies for the one phonon band  of $^{135}$Pr, with the parameters determined as explained in the text, are compared with the experimental excitation energies from the TW2 band \cite{Sen}.
}
\label{Fig.13}
\end{figure}

\begin{figure}[h!]
\includegraphics[width=0.4\textwidth]{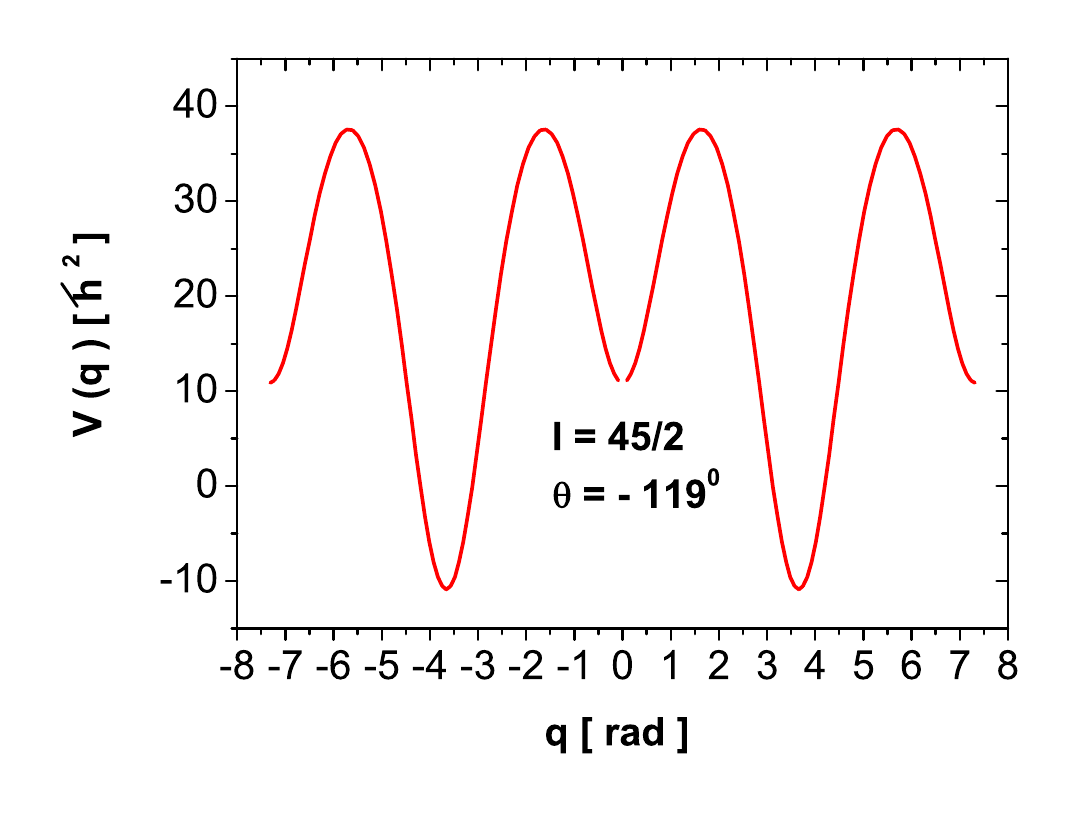}\includegraphics[width=0.4\textwidth]{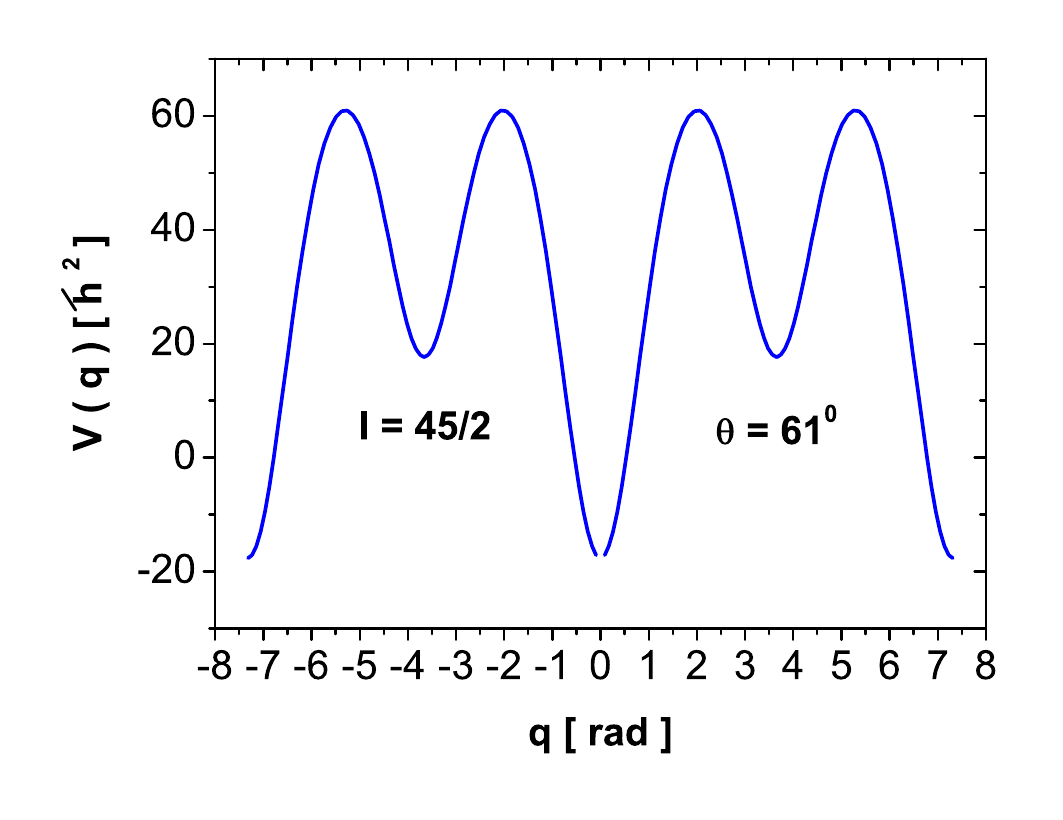}
\caption{(Color online)Potential energies for ${\cal J}_1:{\cal J}_2:{\cal J}_3=91:9:51 [\hbar^2]$ for $\theta=-119^0$ (left panel) and $\theta = 61^0$ (right panel). In both cases j=11/2.
}
\label{Fig.14}
\end{figure}

From Table I we see that the MoI's ordering predicted by our calculations is: ${\cal J}_1>{\cal J}_3>{\cal J}_2$. Due to the adopted fitting procedure this is,  however,  a global result. 
The corresponding potentials $V(\theta)$ and $V(\theta+\pi)$ are shown in Fig.14. From there we remark that $V(\theta)$ has  a deepest minimum in $q=2K$ which contrasts the situation when the maximal MoI is ${\cal J}_2$ where the deepest minimum shows up at $q=0$. This fact asserts that the particle-core coupling may change the MoI's ordering as well as the position of the deepest minimum for the triaxial potential. Such a result represents a specific feature for the present approach.

In order to check whether this ordering holds also when using a different fitting procedure, we fixed the MoI's by equating the calculated excitation energies for the lowest two states of band 1 and the second state of band 2 to the corresponding experimental data, otherwise fixing 
$\theta$  to obtain a global best fit. In this way we found a set of MoI,s which suggests a transverse wobbling regime for the odd system under consideration. Indeed, for $\theta =140^{0}$, the result is ${\cal J}_1=13.53 [\hbar^2/MeV],\;{\cal J}_2=101.76 [\hbar^2/MeV],\;{\cal J}_3=52.94 [\hbar^2/MeV]$. However, the overall agreement with experiment does not improve the results obtained by the former fit method. The potential corresponding to this set of parameters has the deepest minimum at $q=0$  which is consistent with the fact that the maximal MoI is ${\cal J}_2$.
Unfortunately, such a solution is not acceptable since the factor $A$ connecting $H'$ to $H_{rot}$ is negative. Therefore, the minimum of $H'$ is a maximum for $H_{rot}$, which results that the suspected transverse mode is unstable.

Note that being guided by  the scenario where ${\cal J}_2$ is maximum, the fit was performed by using for the wobbling frequency the expression (4.3), obtained by a quadratic expansion of the potential around $q=0$. However, the potential corresponding to the fitted parameters exhibits the deepest minimum in $q=2K$, with the wobbling frequency equal to $\omega^{\prime}$ given by Eq. (4.6). We repeated the fitting  procedure using this frequency in the band definition (7.1). The results are : ${\cal J}_1:{\cal J}_2:{\cal J}_3=89:12:48 [\hbar^2/MeV]$ and $\theta=-71^0$. It is remarkable the fact that the corresponding r.m.s. is equal to that given in Table 1. The reason is that the two frequencies differ from each other by a constant amount, independent of $I$, and therefore they lead to similar excitation energies.

\subsection{ Comment on the chiral features of the wobbling motion}
We recall that a chiral transformation brings a right-handed reference frame to a left-handed one. In the angular momentum space, the change of sign of the a.m. defines a chiral transformation. A system is invariant to a chiral transformation if its rotational energy is preserved when the sense of rotation around an axis is changed.
Note that our starting Hamiltonian is a sum of two terms, one being symmetric  and one antisymmetric with respect to chiral transformations.
\begin{equation}
\hat{H}_{rot}=\hat{H}_s+\hat{H}_a.
\end{equation}
If $|\psi\rangle$ is an eigenstate for $\hat{H}_s$, and $C$ is a chiral transformation, then  $C|\psi\rangle$ is also eigenstate for $\hat{H}_{s}$, and corresponds to the same energy. In this case,
the function $|\psi\rangle$ has the chirality equal to one, since $C|\psi\rangle=|\psi\rangle$.
For $\hat{H}_a$, the above mentioned property changes to : If  $|\psi\rangle$ is an eigenstate of $\hat{H}_{a}$ corresponding to the eigenvalue $E$, then $C|\psi\rangle$ is also eigenstate,
but corresponding to the energy $-E$. Therefore, the eigenvalues of  $\hat{H}_a$ split in two sets, one being the mirror image of the other one. This property is of a chiral nature. The eigenstates of $\hat{H}_a$ have the chirality -1 since $C|\psi\rangle=-|\psi\rangle$. The eigenstates of $\hat{H}_{rot}$ are mixtures of the two chiralities. When there are two sets of energies that are the mirror images of each other, one says that a definite chirality is projected out \cite{Rad016}. In our calculation, the change of $\bf{I}\to-{\bf I}$ is achieved by changing $\theta$ to $\theta  +\pi$. The a.m. dependence of the wobbling frequencies corresponding to $\theta =-119^{0}$, and $\theta =61^{0}$ respectively, is shown in Fig. 15. The look of the potentials  $V$,  and $CVC^{-1}$, are shown in Figs. 14  for $\theta= -119^{0}$, and $\theta= 61^{0}$ respectively, and ${\cal J}_1:{\cal J}_2:{\cal J}_3=91:9:51 \hbar^2MeV^{-1}$. From these two potentials we may extract the symmetric and antisymmetric parts of V.
\begin{equation}
V_s=\frac{1}{2}\left[V(\theta=-119^0)+V(\theta=61^0)\right];\;\;V_a=\frac{1}{2}\left[V(\theta=-119^0)-V(\theta=61^0)\right].
\end{equation}
The two potentials of  definite chirality, are visualized in Fig. 16. A similar analysis can be performed also for the excitation energies. Indeed, Eq.(4.4) expresses explicitly
the dependence of the wobbling frequency on the angle $\theta$, which fixes the orientation of $\bf{j}$. Therefore, it is easy to calculate $\omega_I(\theta+\pi)$, with $\theta =-119^{0}$. The frequencies $\omega_I(\theta)$ and $\omega_I(\theta+\pi)$, with one being the chiral image of the other one, are plotted in Fig.15 for the yrast band. The two curves are parallel to each other, which suggests that the corresponding states have similar properties. However, they do not correspond to states of definite chirality. However, one can extract the symmetric and antisymmetric terms of the excitation energy. Here we give the result for the yrast states:
\begin{eqnarray}
&&E^{exc;1}_{I,s} = A_1 I^ 2 + \left(\omega_{I}(\theta)+ \omega_{I}(\theta+\pi)\right)/2,\nonumber\\
&&I=R+j,\;\;R=0,2,4,..\nonumber\\
&&E^{exc;1}_{I,a} = (2I+1)A_1 j_{1} -IA_2 j_2 + \left(\omega_{I}(\theta)-\omega_{I}(\theta+\pi)\right)/2,\nonumber\\
&&I=R+j,\;\;R=0,2,4,..,\theta=-119^{0}. 
\end{eqnarray} 
Since for asymmetric states, the energies $-E^{exc;1}_{I,a}$ are also eigenvalues of the antisymmetric Hamiltonian, we interpret the two sets of energies $E^{exc;1}_{I,s}$, and $-E^{exc;1}_{I,a}$ 
as defining two bands of chirality +1, and -1, respectively. The relative energies to the head-energy of each band respectively, are plotted in Fig.17. Although the two sets of energies have different dependence on the angular momentum, one is linear (asymmetric) and the other one quadratic in a.m., the energy spacing in the two bands are close to each other, which in fact is a chiral feature of the two bands. Although we don't have enough data to conclude that the  two bands are indeed of real chiral type, due to the above mentioned features, they might be, however, considered as germinos of chiral bands.
Indeed, there are properties unanimously accepted, which prevent us to make a decisive statement on this matter. For example in a chiral band the system rotates around an axis, which doesn't belong to any of the principal planes, while here the rotation axis is a principal axis. However, since in our case the Hamiltonian involves linear terms in the total angular momentum, the wobbling motion and chiral properties seem not  to be disconnected.

\begin{figure}[h!]
\includegraphics[width=0.5\textwidth]{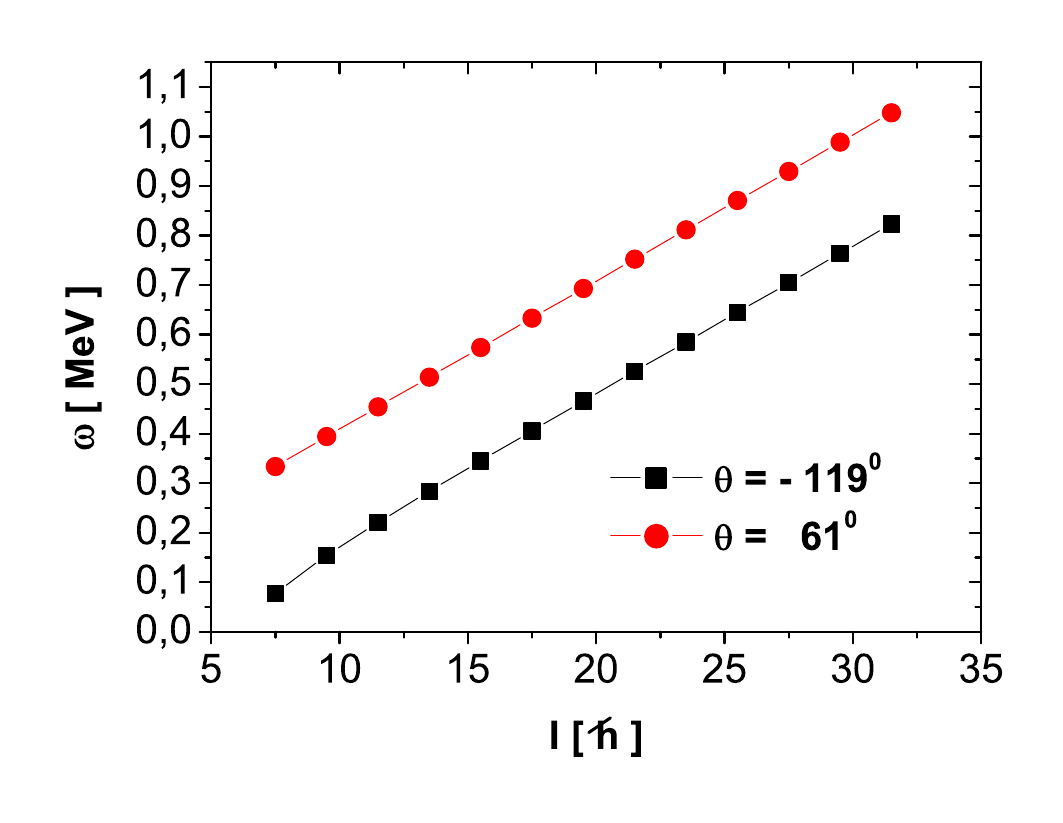}
\caption{(Color online) The wobbling frequencies corresponding to the angles $\theta=-119^0$, and $\theta=61^0$, respectively.
} 
\label{Fig.15}
\end{figure}

\begin{figure}[h!]
\includegraphics[width=0.5\textwidth]{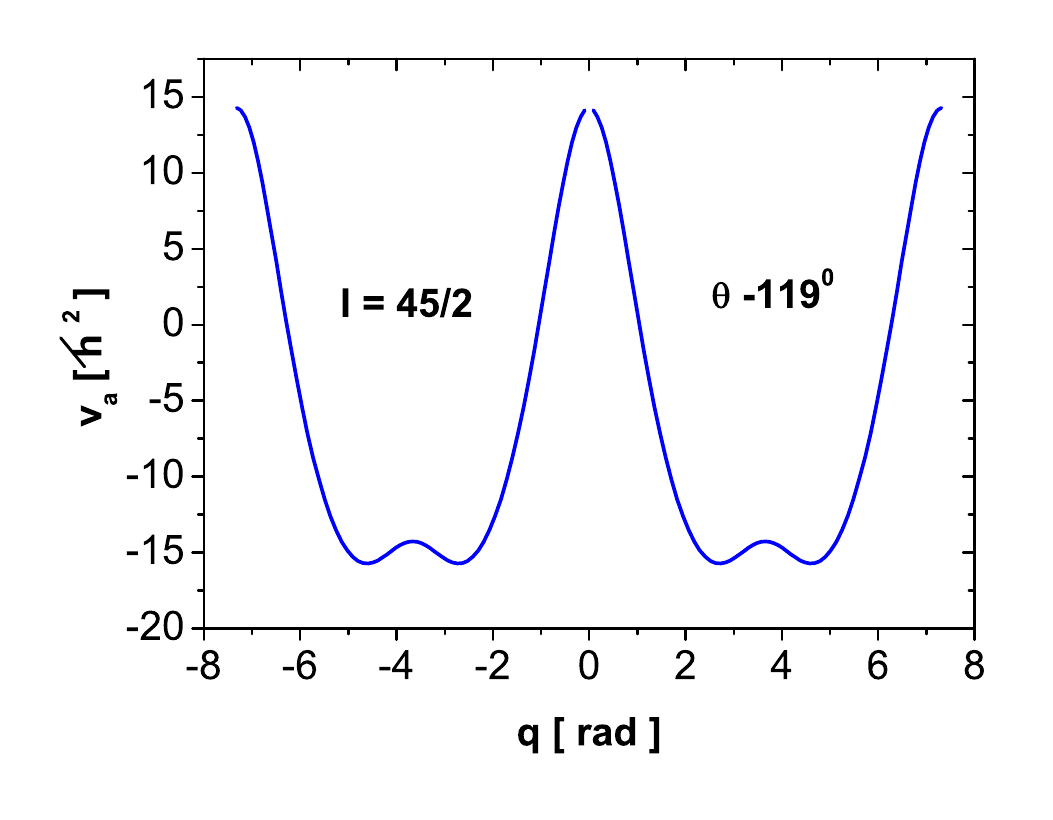}\includegraphics[width=0.5\textwidth]{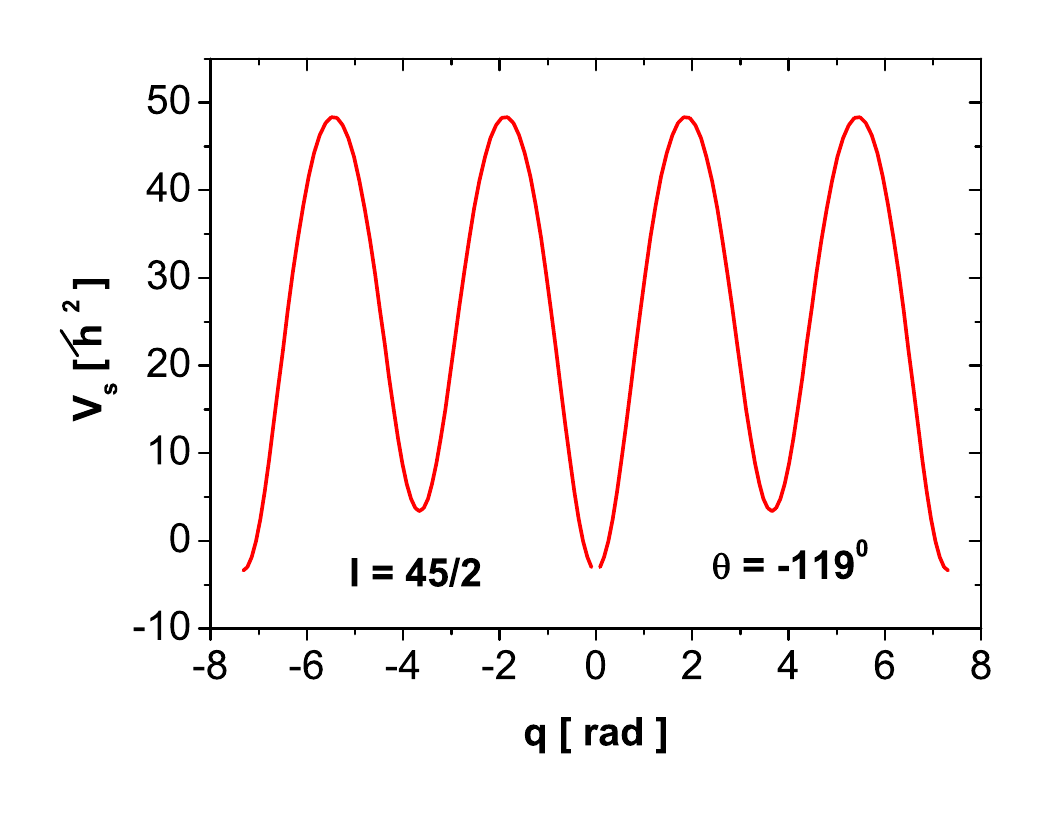}
\caption{(Color online) The antisymmetric (left panel) and symmetric (right) potentials, with respect to the chiral transformations, as function of q.
}
\label{Fig.16}
\end{figure}

\begin{figure}[h!]
\includegraphics[width=0.5\textwidth]{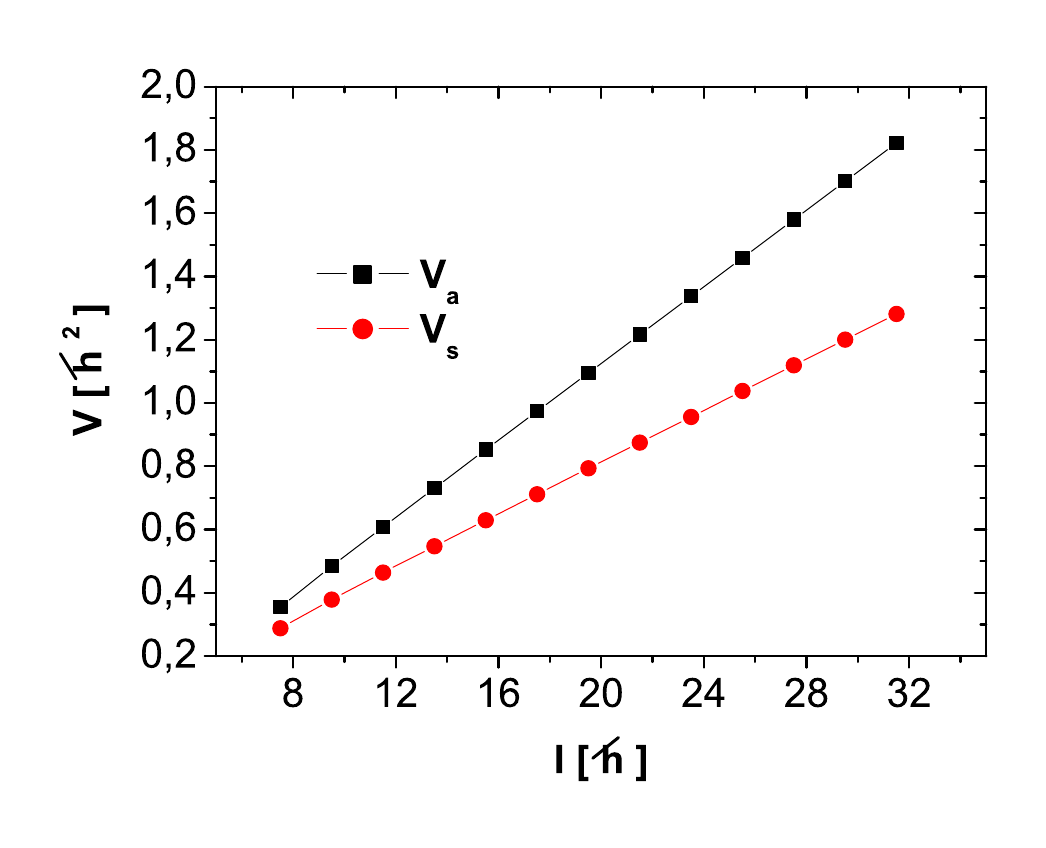}
\caption{(Color online)Energies of symmetric and antisymmetric states against chiral transformations, normalized to the head-energy for each band.
}
\label{Fig.17}
\end{figure}

\clearpage
\subsection{Electromagnetic transitions}
The electric quadrupole reduced transition probabilities were calculated by means of Eq. (6.7), where the reduced matrix elements are those given by (6.5) for the intra-band, and (6.6), for the inter-band transitions. These expressions involve two strength parameters of the quadrupole transition operator, denoted by $\bar{Q}_0$, and $\bar{Q}_2$, respectively. These were determined by fitting two particular branching ratios which results in obtaining the values: $\bar{Q}_{0}=-30.02 \;e.fm^2/\hbar^2$, and $\bar{Q}_{2}=136.39 \;e.fm^2/\hbar^2$.
{\scriptsize
\begin{table}[h!]
\hspace*{-1cm}
\begin{tabular}{|c|cc|cc|cc|}
\hline
&\multicolumn{2}{c|}{$\frac{B(E2;I^-\to (I-1)^-)}{B(E2;I^-\to (I-2)^-)}$} &\multicolumn{2}{c|}{$\frac{B(M1;I^-\to (I-1)^-)}{B(E2;I^-\to (I-2)^-)}$}& \multicolumn{2}{c|}{$\delta_{I^-\to (I-1)^-}$ }\\
& \multicolumn{2}{c|}{ }                            &     \multicolumn{2}{c|}{  $[\frac{\mu_N^2}{e^2b^2}]$}&\multicolumn{2}{c|}{[MeV.fm]}\\
\hline
$I^{\pi}$        & Exp.           &    Th.                                &    Exp.     &Th.                                                       &Exp.     &     Th.  \\
\hline
$\frac{21}{2}^{-}$&0.843$\pm$0.032    &0.400&0.164$\pm$ 0.014&0.08&-1.54$\pm$ 0.09& -1.97\\
$\frac{25}{2}^{-}$&0.500$\pm$0.025    &0.450&0.035$\pm$ 0.009&0.056&-2.384$\pm$0.37& -2.82\\
$\frac{29}{2}^{-}$&$\ge$0.261$\pm$0.014&0.497&$\le$ 0.016$\pm $0.004&0.041&-        & -3.82\\
$\frac{33}{2}^{-}$&-                &0.543&-                        &0.031&-        & -4.96\\
\hline
\end{tabular}
\caption{The calculated branching ratios $B(E2)_{out}/B(E2)_{in}$, and $B(M1)_{out}/B(E2)_{in}$ as well as the mixing ratios $\delta$ are compared with the corresponding experimental data taken from Ref. \cite{Matta}.}
\label{Table II}
\end{table}}
Note that the initial and final wave-functions were determined by the energy calculations, which in fact connect the e.m. transitions to the structure of the model Hamiltonian. 
The results are given in Table II, where the available corresponding experimental data \cite{Matta} are also listed. One notices a reasonable good agreement between the results of our calculations, and the experimental data. The increasing function of the a.m. for the ratio $B(E)_{out}/B(E)_{in}$ is well reproduced. Also, the sign of the mixing ratio, and the increasing behavior with $I$, are consistent with the data, and moreover the results are quite close to the corresponding experimental data.

In principle, to calculate the transition rates $B(E2)$, $B(M1)$ and $E2-M1$ mixing ratios, one needs the $D_2$ symmetric wave function as given in Ref.[1]. Practically, one may use the wave function yielded by
diagonalizing the model Hamiltonian [29], including the coupling of the valence proton to the deformed core. It is worth mentioning that the calculated transition rates in the present paper reproduce reasonably well the data and moreover do not much differ from the results obtained by using the wave functions calculated with the diagonalization procedure [29]. This means that the wave functions used in our approach approximate quite well the exact ones, produced by the diagonalization procedure. In fact this  completes a previous statement [25,26], asserting that the eigenvalues obtained by the diagonalization procedure are identical with the solution of the Schr\"{o}dinger equation in the Bargmann variables. 
\setcounter{equation}{0}
\renewcommand{\theequation}{8.\arabic{equation}}
\section{Conclusions}
The formalism developed in the previous sections may be summarized as follows. A new boson expansion is proposed to describe the wobbling motion in even-odd nuclei. The used Hamiltonian has a simple structure obtained from that describing the even-even core, i.e., a triaxial rotor, by replacing the core angular momentum ${\bf R}$ with ${\bf I-j}$, where ${\bf I}$ and ${\bf j}$ denote the total and odd particle angular momenta, respectively. The coupling term, describing the motion of the odd nucleon in a deformed mean field generated by the core \cite{Davyd} is ignored, since  the odd particle is rigidly coupled to the core, and thereby does not affect the excitation energy spectrum for the odd system. The model Hamiltonian is written in a boson space by using for angular momentum an "elliptic boson expansion". Subsequently, the Bargmann representation is employed, and the eigenvalue equation of the initial model Hamiltonian  is brought to a Schr\"{o}dinger equation form, where the kinetic and potential energy terms are fully separated. The potential is angular momentum dependent, and exhibits several minima, and maxima. Expanding, successively, the potential around the deepest and the local minima, one arrives at two distinct expressions for the wobbling frequency. These results are also obtained within a classical picture, where the phase diagram is constructed for a particular value of I (=15/2). The frequency associated with the deepest minimum is used to describe the energies of the three known wobbling bands in $^{135}$Pr. Due to the presence of linear terms, a possible chiral behavior for the odd system is expected. One succeeds to build up states of a definite chirality. However, it is hard to call the resulting bands as twin bands, although some embryos of them  are present. The electromagnetic reduced transition probabilities are calculated by using for the quadrupole transition operator a quadratic form in the angular momentum. Results for the branching ratios $B(E2)_{out}/ B(E2)_{in}$, and $B(M1)_{out}/ B(E2)_{in}$, as well as for the mixing ratios $\delta$, are compared with the available data. One concludes that the agreement with experimental data for both energies, and e.m. transitions is reasonably good.
It is pointed out that, although we started with the hypothesis that the maximal MoI is that of the 2-axis, the fitting procedure yielded as maximal MoI that of the 1-axis. There is a two fold reason  causing that:a) the renormalization of the MoI, due to the linear terms in angular momentum and b) the Coriolis interaction simulated by the term proportional to $I_1$. At classical level, one showed that the transverse wobbling motion is unstable.

We recall that here the core-${\bf j}$ coupling was neglected. However, the wobbling scheme of the rotor depends on the way how the single particle angular momentum couples to  the rotating core
 [14,29-32]. The present paper shows that the transverse wobbling does not come out in a pure rotor model.

We may assert that the results of the present paper confirm the importance of the boson expansion concept, which was widely used in different contexts of theoretical nuclear physics
\cite{Jol,Rad74,Rad91,Klein91}.

Concluding, the present formalism provides an interesting tool to investigate the theoretical aspects of the wobbling motion in even-odd nuclei and to describe the existent data in a realistic fashion.

{\bf Acknowledgment.} This work was supported by the Romanian Ministry of Research and Innovation through the project PN19060101/2019
\setcounter{equation}{0}
\renewcommand{\theequation}{A.\arabic{equation}}
\section{Appendix A}
Here the a.m. components $\hat{I}_{\pm}$, and $\hat{I}_0$, are written in the Bargmann representation and then the derivative coefficients expanded in the second order around the minimum point of the energy function. The result is:
\begin{eqnarray}
Q_0&=&\frac{1}{\sqrt{6}}\left[3\bar{q}^2\frac{d^2}{d\bar{q}^2} -3(2I-1)\bar {q}\frac{d}{d\bar{q}} 
   +I(2I-1)-I(I-1)(1+k^2)\bar{q}^2\right]\bar{Q}_{0},\\
Q_2&=&\frac{1}{{k^{\prime}}^{2}}\left\{\left[-2+(1+k^2)\bar{q}^2\right]\frac{d^2}{d\bar{q}^2}-(2I-1)(1+k^2)\bar{q}\frac{d}{d\bar{q}}\right.\nonumber\\
   &-&\left.I(1+k^2)+I\left[(I+1)(1+k^2)+k^2(k^2+3)\right]\bar{q}^2\right\}\bar{Q}_{2}.\nonumber
\label{qtran1}
\end{eqnarray}
Note that the magnitude $k (=\sqrt{|u|})$, defined by Eq. (2.10), depends on the angular momentum $I$ due to $u$. Therefore, hereafter, we attach to it a lower index specifying this dependence.
The same procedure is used for he $K+1$, and $K+2$ wave-functions:
\begin{eqnarray}
&&\Phi_{I,-I+1}\equiv\frac{{\cal N}^{(1)}_I}{\sqrt{2I}} \hat{I}_{-}\Phi_{I,-I}=ik_{I}^{'}\frac{{\cal N}^{(1)}_I}{\sqrt{2I}}
\left(-I\bar{q}+\frac{\bar{\omega}_I}{4}\bar{q}^3\right)\Phi_{I,-I},\;\;
\rm{with}\;\;{\cal N}^{(1)}_I=\frac{1}{k_{I}^{\prime}}\sqrt{\frac{\bar{\omega}_I}{I}},\\
&&\Phi_{I,-I+2}\equiv\frac{{\cal N}^{(2)}_I}{\sqrt{4I(2I-1)}} \hat{I}^2_{-}\Phi_{I,-I}=\frac{-{\cal N}^{(2)}_Ik^{\prime}_{I}}{2\sqrt{I(2I-1)}}
\left(-I\bar{q}+\frac{1}{2}\bar{q}^2\frac{d}{d\bar{q}}\right)^2\Phi_{I,-I},\;\rm{with}\; {\cal N}^{(2)}_I=\frac{2\bar{\omega}_I}{k'^{2}_{I}\sqrt{I(2I-1)}}.\nonumber
\end{eqnarray}
The state describing the oscillator vacuum is: 
\begin{equation}
\Phi_{I,-I}=C_Ie^{-\frac{1}{2b_I^2}\bar{q}^2},
\end{equation}
where the norm, and the oscillator length are:
\begin{equation}
C_I=\sqrt{\frac{2}{\pi}}\frac{1}{b_I},\;\;b_I^2=\frac{\hbar}{M\bar{\omega}_I}.
\end{equation}
In our case, the units system is that where $\hbar=1$, while the mass parameter is $M=\frac{1}{2}$. Therefore
\begin{equation}
b_I=\sqrt{\frac{2}{\bar{\omega}_I}},
\end{equation}
where $\bar{\omega}_I=\omega_I/A$, while $\omega_I$ is defined by Eq.(4.2) or by (4.4).
With the above ingredients, the m.e. of the  transition operator corresponding to the intrinsic states $|\Phi_{I,-I}\rangle$ can be evaluated by integration.

\end{document}